\documentclass{aa}  

\usepackage{graphicx}
\usepackage{txfonts}
\usepackage{lipsum}
\usepackage{subcaption}         
                                
\usepackage{lscape}             
                                
\usepackage{placeins}

\usepackage{siunitx}
\newcommand{\kms}{\si{km.s^{-1}}}   
\newcommand{\Htwo}{H$_2$}

\begin{document}

   \title{Spatially resolved optical and mid-infrared spectroscopy of SDSS1335+0728: implications for the origin of the \textit{Ansky} event\thanks{Based on observations collected at the European Southern Observatory under ESO programmes 115.28E2.001 and 115.28E2.002.}}

   \author{P. Sánchez-Sáez\inst{1}\fnmsep\thanks{Corresponding author: paula.sanchezsaez@eso.org}
        \and M. Masterson\inst{2}
        \and L. Hernández-García\inst{3,4}
        \and R. Arcodia\inst{5}
        \and P. Arévalo\inst{6,7}
        \and F. Ávila-Vera\inst{6}
        \and F. E. Bauer\inst{8} 
        \and J. Chakraborty\inst{2}
        \and J. Cuadra\inst{9,7,19}
        \and P. Lira\inst{10,7}
        \and T. Wevers\inst{11}
        \and R. J. Assef\inst{3}
        \and A. Bayo\inst{1}
        \and S. Bernal\inst{10,7}
        \and R. Cartier\inst{12}
        \and Y. Diaz\inst{13}
        \and M. Giustini\inst{14}
        \and H. Guo\inst{15}
        \and D. Ili\'{c}\inst{16,17}
        \and E. Kara\inst{2}
        \and A. B.~Kova\v{c}evi\'{c}\inst{16}
        \and M.L. Martínez-Aldama\inst{18,7}
        \and A. Merloni\inst{19}
        \and G. Miniutti\inst{14}
        \and C. Ricci\inst{20,3}
        \and M. Sniegowska\inst{21} 
        \and G. Calistro Rivera\inst{22}
        \and M. J. Graham\inst{23}
        }
    
   \institute{European Southern Observatory, Karl-Schwarzschild-Strasse 2, 85748 Garching bei München, Germany
   \and Department of Physics \& Kavli Institute for Astrophysics and Space Research, Massachusetts Institute of Technology, Cambridge, MA 02139, USA
   \and Instituto de Estudios Astrof\'isicos, Facultad de Ingenier\'ia y Ciencias, Universidad Diego Portales, Av. Ej\'ercito Libertador 441, Santiago, Chile
   \and Centro Interdisciplinario de Data Science, Facultad de Ingenier\'ia y Ciencias, Universidad Diego Portales, Av. Ej\'ercito Libertador 441, Santiago, Chile
   \and Black Hole Initiative at Harvard University, 20 Garden Street, Cambridge, MA 02138, USA
   \and Instituto de F\'isica y Astronom\'ia, Facultad de Ciencias,Universidad de Valpara\'iso, Gran Breta\~na No. 1111, Playa Ancha, Valpara\'iso, Chile
   \and Millennium Nucleus on Transversal Research and Technology to Explore Supermassive Black Holes (TITANS), 4030000 Concepción, Chile
   \and Instituto de Alta Investigaci{\'{o}}n, Universidad de Tarapac{\'{a}}, Casilla 7D, Arica, 1010069, Chile
   \and Departamento de Ciencias, Facultad de Artes Liberales, Universidad Adolfo Ib\'a\~nez, Av.\ Padre Hurtado 750, Vi\~na del Mar, Chile
   \and Departamento de Astronomía, Universidad de Chile, Camino el Observatorio 1515, Santiago, Chile
   \and Astrophysics \& Space Center, Schmidt Sciences, New York, NY 10011, USA
   \and Centro de Astronom\'ia (CITEVA), Universidad de Antofagasta, Avenida Angamos 601, Antofagasta, Chile
   \and Departamento de Física, Universidad Técnica Federico Santa María, Vicuña Mackenna 3939, San Joaquín, Santiago, Chile
   \and Centro de Astrobiolog\'ia (CAB), CSIC-INTA, Camino Bajo del Castillo s/n, 28692 Villanueva de la Ca\~nada, Madrid, Spain 
   \and Shanghai Astronomical Observatory, Chinese Academy of Sciences, 80 Nandan Road, Shanghai 200030, People's Republic of China
   \and Department of Astronomy, Faculty of Mathematics, University of Belgrade, Studentski trg 16, 11000 Belgrade, Serbia
   \and Hamburger Sternwarte, Universit\"{a}t Hamburg, Gojenbergsweg 112, D-21029 Hamburg, Germany
   \and Astronomy Department, Universidad de Concepción, Casilla 160-C,  4030000, Concepción, Chile
   \and Max-Planck-Institut f\"ur Extraterrestrische Physik, Gie{\ss}enbachstra{\ss}e, D-85748 Garching, Germany
   \and Department of Astronomy, University of Geneva, ch. d’Ecogia 16, 1290, Versoix, Switzerland
   \and Astronomical Institute, Czech Academy of Sciences, Bo\v{c}n\'{i} II 1401, Prague, 14100, Czech Republic
   \and German Aerospace Center (DLR), Institute of Communications and Navigation, Wessling, Germany
   \and California Institute of Technology, 1200 E. California Blvd, Pasadena, CA 91125, USA
   }

   \date{}

\abstract{The galaxy SDSS1335+0728 brightened abruptly in December 2019 (the \textit{Ansky} event) and has since been confirmed as the host of extreme X-ray quasi-periodic eruptions (QPEs) of debated origin. We constrain the origin of its transient activity using spatially resolved VLT/MUSE and JWST MIRI/MRS spectroscopy to characterise the host galaxy and its nuclear accretion history. The stellar kinematics reveal two counter-rotating stellar components and kinematically cold gas ($\sigma_{\rm gas} \lesssim 60$~km~s$^{-1}$), consistent with a past minor merger. The host is dominated by old stars, with ongoing star formation confined to a ring at intermediate radii. Ionisation diagnostics reveal a three-zone structure: a central accretion-powered region, where high-ionisation coronal lines ([\ion{Ne}{VI}], [\ion{Ne}{V}], [\ion{O}{IV}]) are predominantly concentrated, a star-forming ring, and a LINER-like outer region. A Balmer-line light-echo analysis yields a minimum ionising luminosity $\log L_{\rm ion,min} \approx 40.5$~erg~s$^{-1}$ at projected distances beyond the reach of radiation from the \textit{Ansky} event, suggesting nuclear activity prior to 2019. Strong silicate emission at 9.7 and 18\,$\mu$m indicates directly illuminated dust viewed through a low effective mid-infrared (MIR) optical depth; its 9.7\,$\mu$m strength lies at the upper end of the active galactic nuclei (AGN) distribution, closer to that of MIR-selected tidal disruption events (TDEs). The data are consistent with pre-2019 accretion from either a low-luminosity AGN (LLAGN) or a long-lived TDE remnant disc. In both cases, \textit{Ansky} is a slow, faint transient in a $\sim\!10^6\,M_{\odot}$ black hole with ongoing accretion prior to December 2019, offering an alternative to the ``faded AGN'' interpretation proposed for some QPE hosts.}

   \keywords{Galaxies: nuclei --  Galaxies: active -- Galaxies: individual : SDSS1335+0728 -- X-rays: individuals: Ansky
               }

\titlerunning{Spatially resolved optical and mid-infrared spectroscopy of \textit{Ansky}'s host galaxy}
\authorrunning{Sánchez-Sáez et al.}

\maketitle
\nolinenumbers

\section{Introduction}\label{intro}

The galaxy SDSS~J133519.91+072807.4 (hereafter SDSS1335+0728, $z=0.024$), which had shown no optical variations for at least two decades, exhibited a sudden optical brightening in December 2019, detected by the \textit{Zwicky} Transient Facility (ZTF; \citealt{Bellm19}), an event designated as \textit{Ansky} (from its ZTF object ID ZTF19acnskyy). \cite{Sanchez-Saez24} presented the discovery of the nuclear activity in SDSS1335+0728, and proposed that the variations observed could be either explained by a $\sim 10^6 M_{\odot}$ active galactic nucleus (AGN) that was turning on (an awakening AGN; \citealt{Arevalo24}) or by an exotic tidal disruption event (TDE; the disruption of a star when it passes very near to a black hole, BH, with a mass of $\lesssim 10^8  M_{\odot}$; \citealt{Rees88,vanVelzen20}), due to its flat, slow optical decay and persistent variability. \cite{Sanchez-Saez24} also reported the evolution of the [OIII]$\lambda$5007$\AA$ line, which increased its flux $\sim 3.6$ years after the first ZTF alert, and a delayed mid-infrared (MIR) echo, observed in 2022, implying an unusually large dusty inner radius ($>1.2$~pc) for a $\sim10^6 M_\odot$ BH. A recent work by \cite{Zhu25} presented ultraviolet (UV) spectroscopy from the Hubble Space Telescope (HST), which revealed a featureless, steep continuum (power-law index of -2.6) and a lack of broad emission lines. \cite{Zhu25} concluded that the \textit{Ansky} event could be explained by a low-luminosity, slowly evolving, featureless TDE produced by the disruption of a post-main-sequence star. 

\textit{Ansky} is most remarkable for its X-ray properties. \cite{Hernandez-Garcia25a} reported that since February 2024, extreme X-ray Quasi-Periodic Eruptions (QPEs; recurring soft X-ray transients associated with super massive BHs, SMBHs; \citealt{Miniutti19,Giustini20,Arcodia21,Arcodia24a,Arcodia25,Chakraborty21,Chakraborty25a,Quintin23,Nicholl24,Hernandez-Garcia25a,Baldini26}) were associated with the \textit{Ansky} event, with the highest fluxes, longest timescales ($\sim$1.5-day duration), and largest energy outputs ($\sim10^{48}$ ergs per burst) of any known QPE source, recurring approximately every 4.5 days. The physical mechanism driving QPEs remains an open question. Proposed models generally fall into two broad categories: accretion disc instabilities \citep{Raj21,Pan22,Pan23,Sniegowska23,Kaur23} and orbital interactions between the central SMBH and/or its disc and a secondary body in an extreme mass-ratio inspiral (EMRI) configuration. The latter scenario can be further divided into models invoking mass transfer at pericenter from the companion star (e.g., \citealt{King20,King22,Lu23,Olejak25}), and models that invoke collisions between a stellar-mass companion and a SMBH accretion disk (e.g., \citealt{Franchini23,Linial23b,Linial25,Zhou24a}).

Soon after the discovery of \textit{Ansky}'s QPEs, \cite{Chakraborty25b} presented evidence of a time-evolving P Cygni profile, possibly emerging from the debris ejected by the orbiter-disk collision, in the context of EMRI-driven QPEs. Later, in 2025, high-cadence monitoring revealed a dramatic evolution: the recurrence time doubled to $\sim$10 days, flare durations extended to up to 4 days, and the eruptions became four times more energetic \citep{Hernandez25b}. Recent X-ray observations, obtained between February 2025 and January 2026, yielded the first direct measurement of a period derivative in a QPE \citep{Chakraborty26A}, revealing that Ansky's recurrence period has smoothly increased at a rate of $\dot{P}\approx (1.7\pm 0.02)\times10^{-2}$ d d$^{-1}$. This positive derivative poses a unique challenge to standard EMRI models, which typically predict a shrinking orbit. Furthermore, recent monitoring has provided the first-ever detection of recurrent UV variability temporally coupled to the X-ray QPEs \citep{Guo26}. These coherent UV modulations lag the X-ray eruptions by approximately 1 day, a discovery likely facilitated by Ansky's unusually long recurrence timescale, which prevents the temporal smearing of the UV response seen in more rapid QPEs, or the larger accretion disc of \textit{Ansky} respect to other QPEs \citep{Hernandez-Garcia25a}.

While the unprecedented multiwavelength monitoring of \textit{Ansky} provides a real-time view of its activity, understanding its current and past accretion history requires spatially resolved spectroscopy of SDSS1335+0728's extended environment. Here, optical and infrared Integral Field Spectroscopy (IFS) observations of other QPE host galaxies provide an essential reference frame.

IFS observations, primarily utilizing the Multi Unit Spectroscopic Explorer (MUSE; \citealt{muse}) at the Very Large Telescope (VLT), have provided crucial insights into the environments that host QPEs. Observations presented by \cite{Wevers24c} reveal that three out of five known QPE host galaxies observed with MUSE (GSN 069, RX J1301, and eRO-QPE2) present Extended Emission-Line Regions (EELRs) of ionised gas that extend up to 10 kpc from the galactic nucleus, while \cite{Xiong25} presented similar observations for the host of AT2019qiz. These EELRs exhibit peculiar kinematics: they are physically and kinematically decoupled from the stellar continuum, displaying low velocities ($<100$ km/s) and low velocity dispersions ($\lesssim 75$ km/s). Spectral line diagnostics indicate that the gas is being photoionised by a hard, nonstellar continuum. However, the current bolometric luminosities of these galactic nuclei are insufficient to power the observed large-scale emission. From this energy deficit, recent works have proposed that these galaxies hosted an AGN that recently faded (within the last $\sim15,000$ to $30,000$ years; e.g., \citealt{Jiang25,Xiong25}). Conversely, recent theoretical work by \cite{Mummery25} proposes a compelling alternative (in agreement with observations presented in \citealt{Wevers24b,Wevers24c}): the long-lived accretion discs formed by TDEs themselves produce sufficient ionising radiation to power EELRs on galaxy-wide scales (extending up to $\sim10^4$ light-years). This mechanism naturally explains the observations without requiring a faded AGN: galaxies with intrinsically high TDE rates are simply more likely to exhibit these macroscopic EELRs, which act as light echoes from prior stellar disruptions. Furthermore, QPE host galaxies show a striking statistical resemblance to TDE host galaxies \citep{Wevers24b,Wevers24c}. Both are significantly overrepresented in post-starburst (PSB) and quiescent Balmer-strong (QBS) galaxies, indicating a shared preference for gas-rich, post-merger environments where a massive BH has recently ceased its active accretion phase. This overlapping host demographic suggests that TDEs and QPEs share a common formation channel, potentially linked to the disruption of nuclear stellar dynamics following a minor merger.

Recent MIR observations by the James Webb Space Telescope (JWST) MIRI Medium-Resolution Spectrometer (MIRI/MRS; \citealt{miri}) have revolutionized our understanding of the circumnuclear dust and gas around TDEs. JWST observations of MIR-selected TDEs \citep{Masterson25} reveal striking structural properties. These transient events produce high ionisation potential (IP) lines that are consistent with ionisation by an accretion-driven hard continuum. Furthermore, because TDEs fundamentally lack the long-lived accretion flow required to maintain a thick, obscuring dusty torus \citep{Wada2012}, their preexisting circumnuclear dust is directly illuminated by the transient flare. \citet{Masterson25} found this dust to be well described by highly optically thin models, producing massive, broad silicate emission features around 9.7 $\mu$m and 18 $\mu$m that are significantly stronger than those of all AGN in the comparison samples they considered. However, this silicate emission is not unique to TDEs. Recent MIRI/MRS observations of low-luminosity AGNs (LLAGNs) have revealed that LLAGNs also exhibit broad, unresolved silicate emission at 9.7 $\mu$m and 18 $\mu$m on parsec scales \citep{Goold26}, although these features are comparatively weaker than those seen in TDEs. LLAGNs also produce rich emission-line spectra with high-IP coronal lines like [\ion{Ne}{V}]$\lambda14.32\mu$m and [\ion{O}{IV}]$\lambda25.89\mu$m. However, unlike typical active accretion flows, LLAGNs are believed to be sustained by low-density radiatively inefficient accretion flows \citep[RIAFs;][]{Narayan1994}, with low Eddington ratios ($L_{\text{Bol}}/L_{\text{Edd}}<10^{-3.5}$), which produce relatively less [Ne V] emission because their truncated accretion disks lack the abundant high-energy UV and soft X-ray photons required to efficiently excite it. 

In this work, we present optical and MIR IFS observations of SDSS1335+0728, obtained using the VLT/MUSE and the JWST MIRI/MRS instruments. We used these observations to characterize \textit{Ansky}'s host galaxy and, from this, better understand the history of SDSS1335+0728 before the 2019 transient event. The paper is organised as follows. In Section \ref{data}, we present the VLT/MUSE and JWST MIRI/MRS observations. In Section \ref{results}, we present the data analysis performed to understand SDSS1335+0728's kinematics and past and recent evolution. In section \ref{discussion}, we discuss the implications of the results presented for the evolution of SDSS1335+0728, and the origin of \textit{Ansky}. Finally, in Section \ref{conclusions}, we conclude and summarise our findings. We adopt the cosmological parameters $H_0=70$ km s$^{-1}$ Mpc$^{-1}$, $\Omega_m=0.3$ and $\Omega_\Lambda=0.7$. 

\section{Data}\label{data}

\subsection{VLT/MUSE data reduction and spectral fitting}\label{muse_data}

VLT/MUSE observations (ESO programme 115.28E2, PI: Hernández-García) were obtained using the wide (1\arcmin$\times$1\arcmin field of view, FoV; $0\farcs2$~pixel$^{-1}$ pixel scale) and narrow (7\farcs5$\times$7\farcs5 FoV; $0\farcs025$~pixel$^{-1}$ pixel scale) field modes (WFM and NFM), between 2025-05-03 and 2025-07-15. In total, we obtained two 1-hour observing blocks (OBs) with the WFM and four 1-hour OBs with the NFM. The data were reduced using the MUSE instrument pipeline \citep{muse_pipeline}, and stacked OBs were obtained for each MUSE mode from the ESO MUSE-DEEP data products\footnote{\url{https://www.eso.org/rm/api/v1/public/releaseDescriptions/102}}. 

\begin{figure}[htbp]
\begin{center}
        \includegraphics[width=0.85\linewidth]{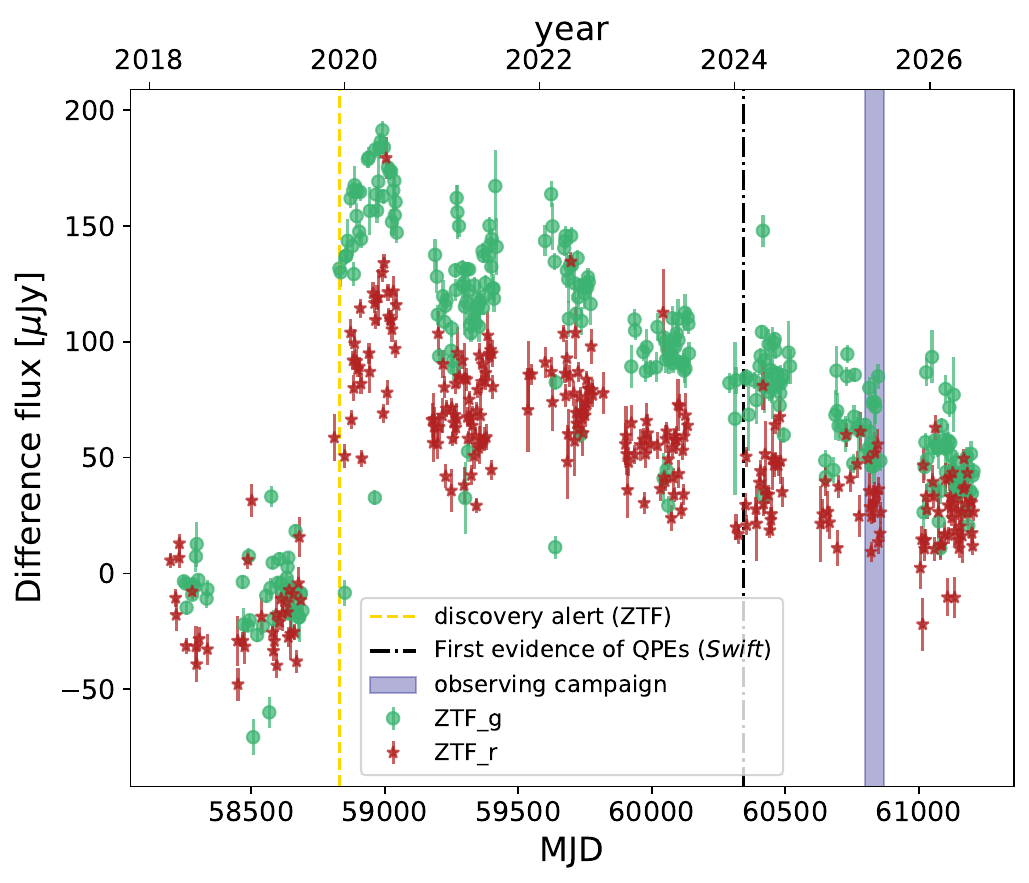}

    \caption{ZTF $g$ (green) and $r$ (red) band forced photometry light curve of \textit{Ansky}. A one-day binning is applied to reduce the noise typically observed in ZTF forced photometry light curves. The orange dashed line shows the first alert Modified Julian Day (MJD), the black dashed-dotted line shows the first evidence of X-ray QPEs (from \textit{Swift} observations), and the blue region shows the window encompassing the VLT/MUSE and JWST MIRI/MRS observations. The light curve shows that \textit{Ansky}'s flux has not yet reached the values observed prior to December 2019, but it is slowly approaching them. }
    \label{fig:ztf_lc}
\end{center}

\end{figure}

For the WFM data, a spatial sub-region of $90\times90$ spaxels centred on the source was extracted prior to analysis. All wavelengths were converted to the rest-frame using the source redshift $z=0.024$, and the spectral range was restricted to $4600$--$7200$~\AA{} (rest-frame) to encompass the principal optical emission lines whilst excluding poorly calibrated regions at the cube edges. 

Before fitting, a foreground Milky Way (MW) extinction correction was applied to each spectrum. The colour excess $E(B-V)_{\rm MW}$ was retrieved from the \cite{Schlegel98} dust map via a query to the NASA/IPAC IRSA Dust service at the cube pointing coordinates. Fluxes were corrected using the \citet[CCM89]{Cardelli89} extinction law with $R_V = 3.1$, evaluated at observed-frame wavelengths.

The nuclear position (SMBH position) was identified as the peak of the white-light (wavelength-collapsed) image. The signal-to-noise ratio (SNR) per spaxel was computed in a rest-frame continuum window of $5050$--$5150$~\AA, defined as the ratio of the median signal to the square root of the median variance across that window. Spatial binning was performed with the PowerBin\footnote{\url{https://pypi.org/project/powerbin/}} adaptive algorithm \citep{Cappellari2025}, targeting a minimum SNR of 20 per bin. Voronoi bins exceeding 400 spaxels (NFM) or 100 spaxels (WFM) were discarded to avoid excessively large bins in the low-surface-brightness outskirts. The spectrum of each spatial bin was formed by inverse-variance weighting of the contributing spaxels, with the MW extinction correction applied multiplicatively as a function of wavelength.

Stellar kinematics and emission-line fluxes were extracted simultaneously from each spatial bin using the Penalised Pixel-Fitting code \cite[\texttt{pPXF};][]{Cappellari2023}, fitting MILES stellar population synthesis (SSP) templates \citep{Vazdekis2010} over the rest-frame range $4600$--$7200$~\AA, with templates convolved to the wavelength-dependent MUSE line spread function (LSF) \citep{Bacon2017} prior to fitting. In the nuclear region (circular region with radius $r < 0\farcs5$ for NFM, and $r < 1\farcs0$ for WFM), an additional accretion-driven power-law continuum component was included to account for the ongoing accretion activity confirmed by contemporaneous ZTF photometry (see Figure \ref{fig:ztf_lc}). A description of the fitting configuration, LSF treatment, and emission-line template setup is provided in Appendix~\ref{app:ppxf}.

The point spread function (PSF) of the WFM observations was estimated empirically from unsaturated field stars detected within the MUSE FoV, and independently validated against the spatial profile of the accretion-driven power-law continuum component recovered by \texttt{pPXF}, finding consistent results. For the NFM, no field stars are available, and thus the PSF FWHM was estimated directly from the accretion-driven power-law continuum profile. Full details of both methods are given in Appendix~\ref{app:psf}. Hereafter, we use the PSF derived from the field stars for the WFM ($\text{PSF}_{\text{WFM}}=0\farcs62$), and from the accretion-driven power-law continuum profile for the NFM ($\text{PSF}_{\text{NFM}}=0\farcs19$).

At the redshift of SDSS1335+0728, 1\arcsec\ corresponds to $\approx0.48$~kpc, so the analysed WFM region ($18\arcsec\times18\arcsec$) spans $\approx8.7\times8.7$~kpc and the NFM FoV $\approx3.6\times3.6$~kpc, with PSF FWHMs of $\approx0.30$~kpc (WFM) and $\approx0.09$~kpc (NFM).

\subsection{JWST MIRI/MRS data reduction and spectral fitting}\label{jwst_data}

JWST MIRI/MRS observations (JWST Cycle~4; Program ID: 8245; PI: P.~S\'anchez-S\'aez) were obtained in July 2025. We obtained a single observation with a 4-point point-source dither pattern, and $100~\mathrm{groups} \times 1, 1, 2$ integrations in the short, medium, and long grating settings, respectively, using the full-frame fast readout mode. The galaxy emission fills the entire MIRI/MRS FoV, and thus, we used an offset pointing for background estimation. MIRI/MRS provides spectra from 4.9--27.9\,$\mu$m across four channels (channel~1: 4.9--7.65\,$\mu$m; channel~2: 7.51-11.70\,$\mu$m; channel~3: 11.55--17.98\,$\mu$m; channel~4: 17.70--27.9\,$\mu$m). The FoV and default cube spaxel size scale with channel, ranging from $3\farcs2\times3\farcs7$ and 0\farcs13\,pixel$^{-1}$ in channel~1 to $6\farcs6\times7\farcs7$ and 0\farcs35\,pixel$^{-1}$ in channel~4, while the PSF FWHM scales with wavelength as $\mathrm{FWHM}\,[\mathrm{arcsec}] = 0.033\,\lambda_{\rm obs}\,[\mu\mathrm{m}] + 0.106$, ranging from $\approx0\farcs27$ at 4.9\,$\mu$m to $\approx1\farcs03$ at 27.9\,$\mu$m. At the redshift of SDSS1335+0728, these correspond to FoVs of $\approx1.5\times1.8$~kpc (channel~1) to $\approx3.2\times3.7$~kpc (channel~4), spaxel sizes of $\approx63$--$170$~pc, and PSF FWHMs of $\approx0.13$--$0.50$~kpc.

We reduced the data using the JWST Science Calibration Pipeline \citep[version~1.18.1;][]{Bushouse2024} with Calibration Reference Data System (CRDS) context \texttt{jwst\_1364.pmap}, following the standard three-stage procedure, as in \citet[see their Section~2 for further details]{Masterson25}. This includes master background subtraction using the dedicated offset observation and a residual fringing correction applied to the extracted 1D spectrum. The dithered exposures were combined into twelve flux-calibrated spectral cubes, and the nuclear 1D spectrum was extracted using the \texttt{extract\_1d} step of the \texttt{Spec3} pipeline assuming a point source geometry. This step extracts a 1D spectrum with a conical aperture of radius equal to one PSF FWHM and center selected with the automatic centroiding algorithm. It also applies an aperture correction that accounts for light lost outside of the central aperture and subtracts off background emission from an annular region, for which we use the default parameters of the \texttt{extract\_1d} step. The resulting 12 sub-band spectra were combined into a single continuous spectrum by averaging the flux in the overlapping wavelength regions between adjacent sub-bands.

Emission-line fluxes, velocity centroids, and line widths were measured from the stitched 1D spectrum by fitting a single Gaussian profile plus a local linear continuum independently to a set of targeted transitions spanning high-IP coronal lines, low-IP ISM tracers, and \Htwo\ rotational lines (see Appendix~\ref{app:jwst_linefitting} for details). Lines were considered detected at SNR\,$\geq 3$, $|\Delta v| \leq 1000$\,\kms, and FWHM\,$\geq 100$\,\kms. Spatially resolved maps of line flux were derived from the spectral cubes via spectral moment analysis within a $\pm600$\,\kms\ window after local continuum subtraction, with an independent per-spaxel Gaussian fit applied where the peak SNR exceeded 3; all maps were reprojected onto a common 0.20\,arcsec\,pixel$^{-1}$ grid and masked below SNR\,$= 3$.

\begin{figure*}[htb!]
\begin{center}
        \includegraphics[width=\linewidth]{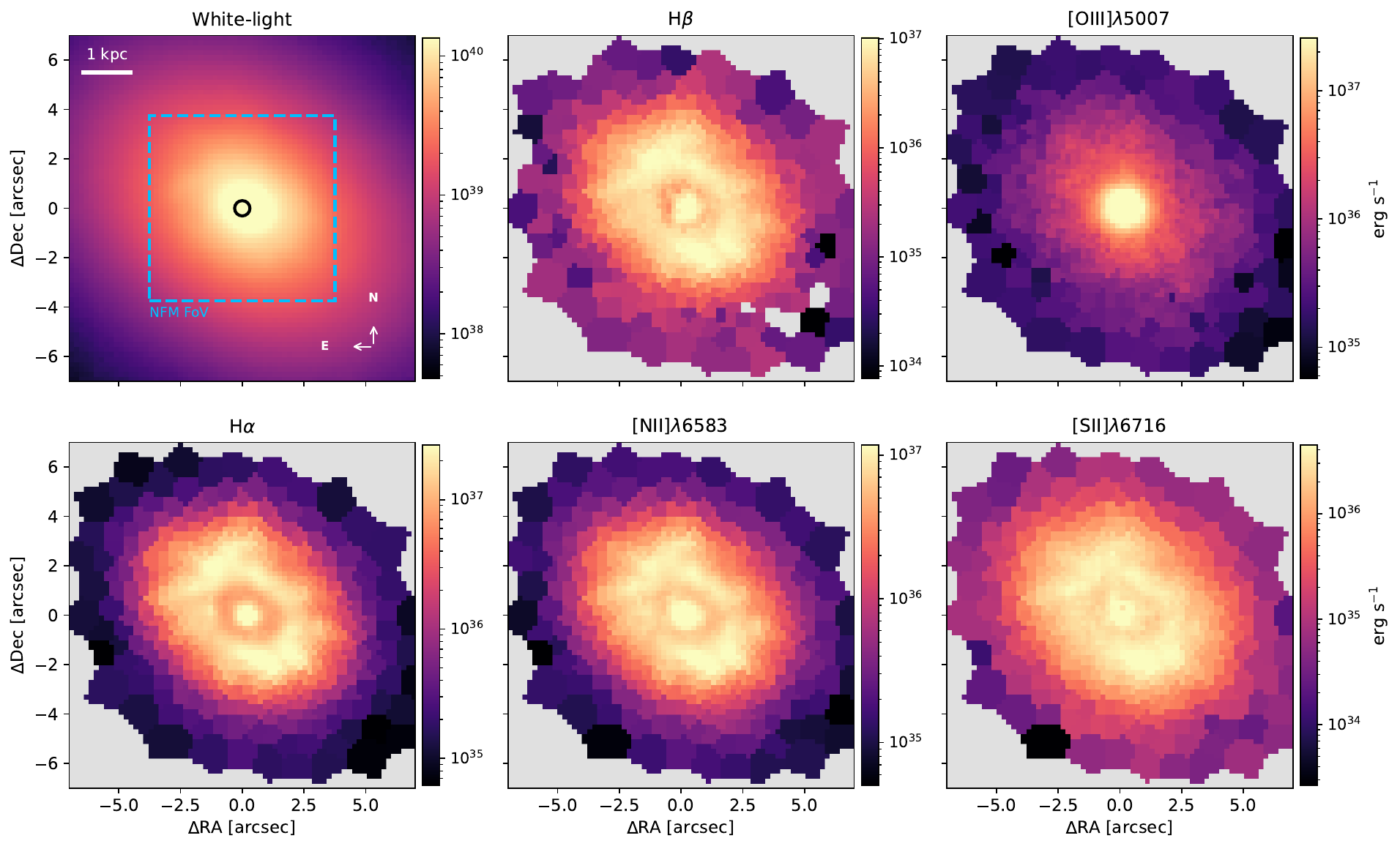}

    \caption{Spatially resolved emission-line luminosity maps (in units of erg~s$^{-1}$) of SDSS1335+0728 from the MUSE WFM datacube, displayed on a logarithmic colour scale. Top row, left to right: white-light continuum image, H$\beta$, and [\ion{O}{III}]$\lambda$5007. Bottom row: H$\alpha$, [\ion{N}{II}]$\lambda$6583, and [\ion{S}{II}]$\lambda$6716. The white image panel shows the PSF FWHM (black circle), the orientation of all the maps presented in this work (white arrows), and as reference, the size of the MUSE NFM FoV (cyan square). All maps are centred on the AGN position.
\label{fig:muse_flux_maps}}
\end{center}
\end{figure*}

\section{Results}\label{results}

In the following sections, we present the results obtained from the spectral analysis of both VLT/MUSE and JWST MIRI/MRS observations. MUSE is used to characterise the kinematics, stellar populations, optical ionisation structure, and light echo (Sect.~\ref{muse_results}). MIRI/MRS provides the nuclear MIR spectrum and the MIR line and ionisation maps (Sect.~\ref{jwst_results}).

\subsection{VLT/MUSE data cube analysis}\label{muse_results}

\subsubsection{Spatially resolved emission-line morphology}
\label{sec:muse_fluxmaps}

Figure~\ref{fig:muse_flux_maps} presents the emission-line flux maps extracted from the WFM datacube for H$\beta$, [\ion{O}{III}]$\lambda$5007, H$\alpha$, [\ion{N}{II}]$\lambda$6583, and [\ion{S}{II}]$\lambda$6716, together with the white-light (wavelength-collapsed) image. All maps are shown over a $14\arcsec \times 14\arcsec$ FoV centred on the galactic nucleus, with coordinates expressed as offsets in right ascension and declination from the nuclear position.

The emission-line maps show spatially extended nebular emission well beyond the nuclear region, reaching projected distances of $\sim 6$--$7\arcsec$ ($\sim 3.4$~kpc) from the nucleus. The [\ion{O}{III}]$\lambda$5007 emission is the most centrally concentrated of the mapped lines, peaking sharply at the nucleus, consistent with recent ionisation by a compact central source. In addition, fainter [\ion{O}{III}]$\lambda$5007 emission extends to $\sim 6\arcsec$, forming a roughly symmetric structure around the nucleus consistent with an EELR. In contrast, H$\alpha$, H$\beta$, [\ion{N}{II}]$\lambda$6583, and [\ion{S}{II}]$\lambda$6716 show broadly similar morphologies, with enhanced emission at the nucleus but also in a ring at 1\arcsec--3\arcsec of the nucleus. The PSF FWHM of the WFM observations ($\mathrm{FWHM}_{\rm PSF} = 0\farcs62$) is indicated in the first panel as a black circle, confirming that the nebular emission is spatially resolved on scales significantly larger than the PSF.

In section \ref{sec:results_cumflux} of the Appendix, we show the cumulative [\ion{O}{III}]$\lambda$5007 luminosity as a function of aperture radius. The measured luminosity within 1\arcsec, above the values reported in \cite{Sanchez-Saez24}, confirms that the evolution proposed in that work was real, and not due to aperture effects.

\subsubsection{Emission-line diagnostic classifications}
\label{sec:results_bpt}

\begin{figure*}[htbp]
\begin{center}
        \includegraphics[width=\linewidth]{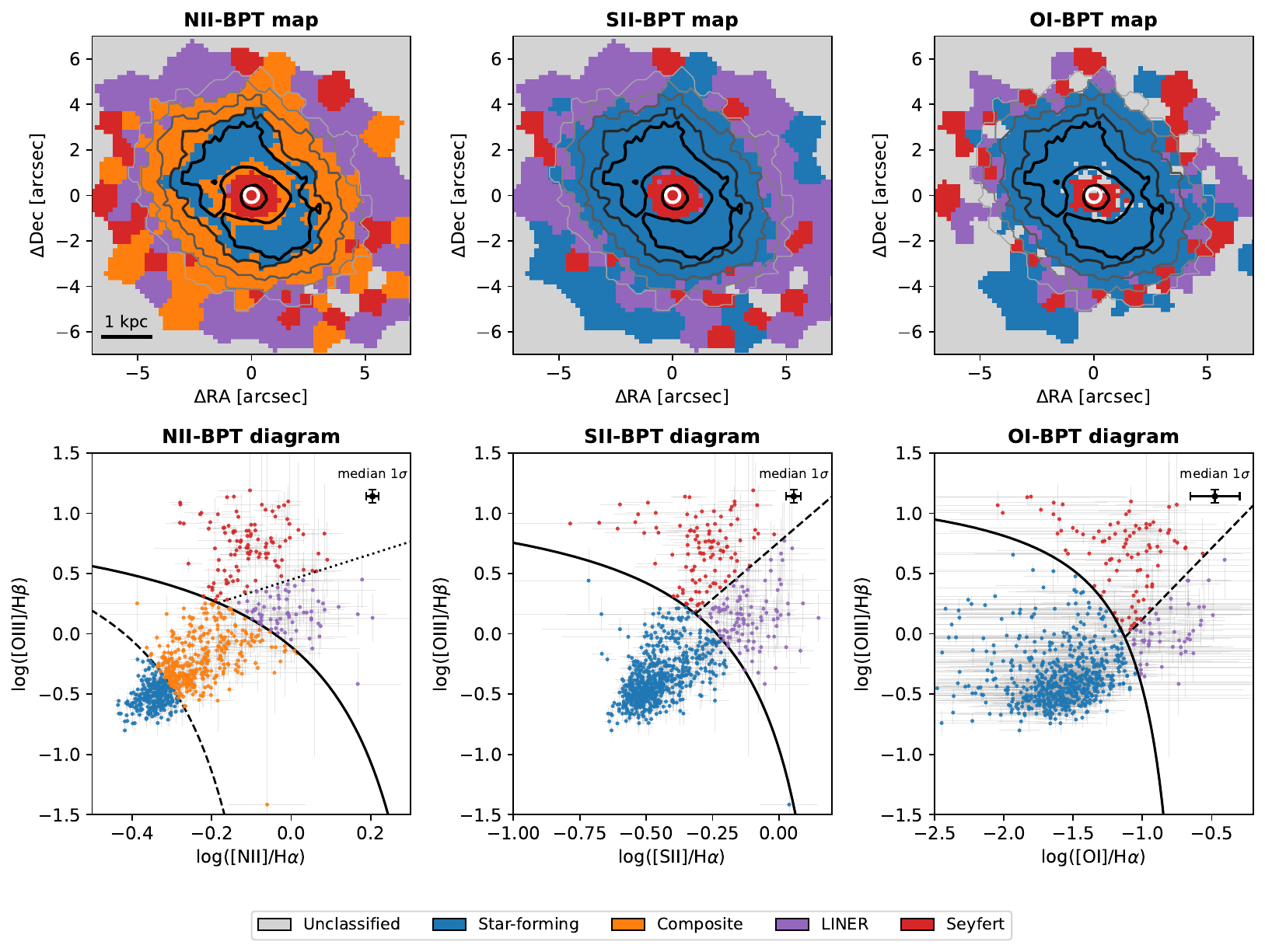}

 \caption{BPT diagnostic classification of the MUSE WFM field. Top row: spatially resolved classification maps for the [\ion{N}{II}]- (left), [\ion{S}{II}]- (centre), and [\ion{O}{I}]-BPT (right) diagnostics, colour-coded by excitation class as indicated in the legend. Black contours show the H$\alpha$ emission (percentiles 50\% to 90\%), with the brightest level highlighted in a darker colour. The derived PSF size is shown with a white circle. Bottom row: corresponding diagnostic diagrams with the \citet{2001ApJ...556..121K} (solid) and \citet{Kauffmann03} (dashed; [\ion{N}{II}] only) demarcation lines. Grey error bars show the $1\sigma$ uncertainties of individual bins; the black error bar marks the median uncertainty.
\label{fig:bpt_wfm}}
\end{center}
\end{figure*}

\begin{figure*}[htbp]
\begin{center}
\includegraphics[width=\linewidth]{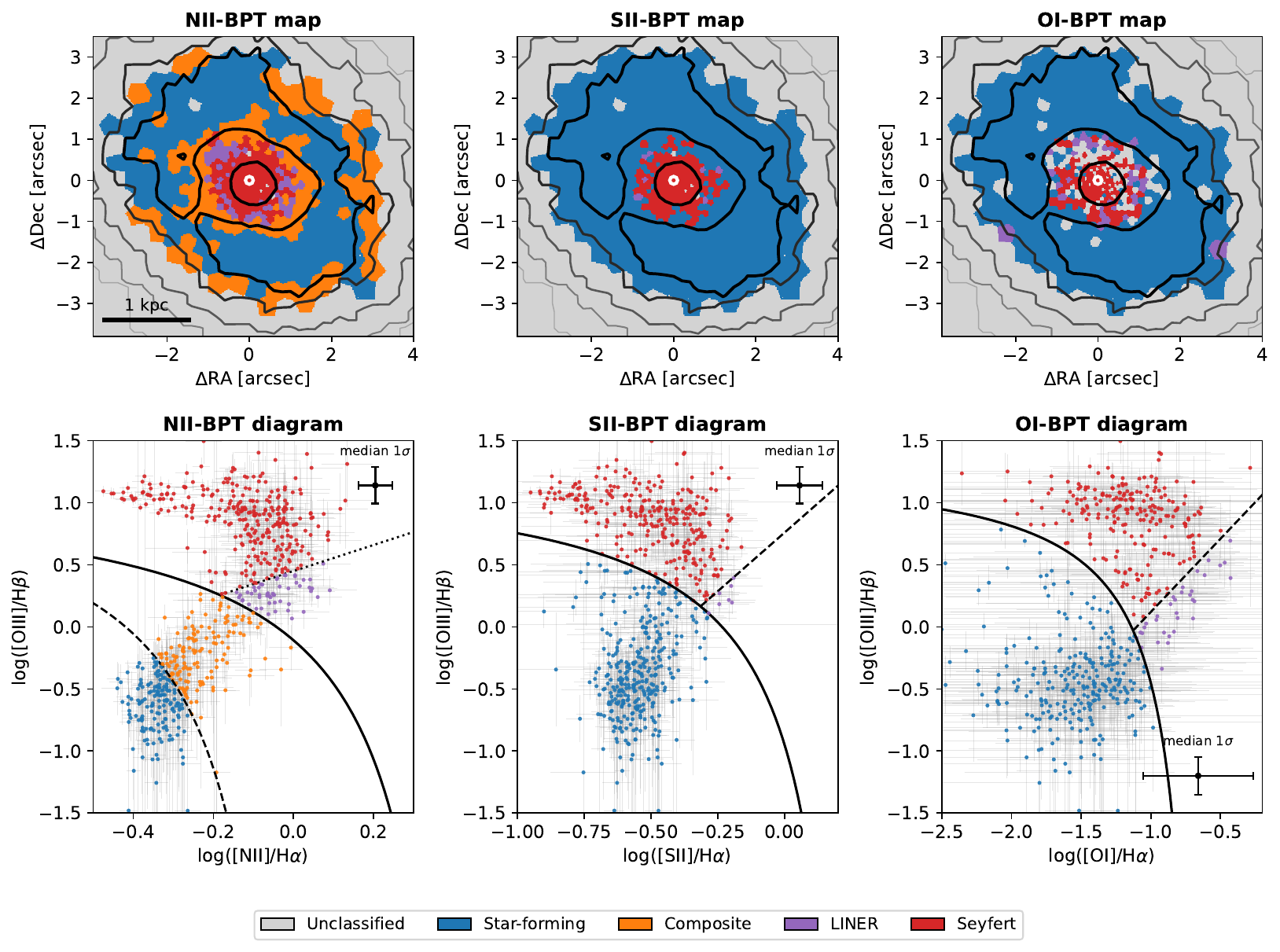}

\caption{As Figure~\ref{fig:bpt_wfm}, but for the MUSE NFM datacube ($\mathrm{FWHM}_{\rm PSF} = 0\farcs19$). The H$\alpha$ contours are derived from the WFM data, as in Figure~\ref{fig:bpt_wfm}, so that the same structures can be traced in both fields.
\label{fig:bpt_nfm}}
\end{center}
\end{figure*}

Figure~\ref{fig:bpt_wfm} presents the BPT diagnostic (after Baldwin, Phillips, and Terlevich; \citealt{Baldwin81}) classification of the WFM field, and Figure~\ref{fig:bpt_nfm} the corresponding NFM classification. The BPT diagrams use ratios of strong optical emission lines, [\ion{O}{III}]$\lambda$5007/H$\beta$ versus [\ion{N}{II}]$\lambda$6583/H$\alpha$, [\ion{S}{II}]$\lambda\lambda$6716,6731/H$\alpha$, or [\ion{O}{I}]$\lambda$6300/H$\alpha$, to distinguish the dominant ionisation mechanism in each spatial bin: photoionisation by young stars (star-forming), a harder radiation field from an AGN or accretion-driven emission (Seyfert), a mixture of both (composite), or low-ionisation emission line region (LINER), separated by the theoretical and empirical demarcation curves of \citet{2001ApJ...556..121K} and \citet{Kauffmann03}. Each figure shows the spatial classification maps (top row) with the corresponding diagnostic diagrams (bottom row) for the [\ion{N}{II}]-, [\ion{S}{II}]-, and [\ion{O}{I}]-BPT planes. As described in Section \ref{muse_data}, spaxels were spatially binned with PowerBin to a target continuum SNR of 20 in the rest-frame 5050--5150~\AA\ window, and the classification is performed per bin. In both Figures \ref{fig:bpt_wfm} and \ref{fig:bpt_nfm}, bins are colour-coded by their BPT class. Bins without a BPT classification (shown in grey) correspond either to regions excluded from the spatial binning, or to bins where at least one of the lines considered is not detected, i.e.\ its flux is set to zero by the non-negative pPXF fit. The $1\sigma$ uncertainties of the line ratios, shown in the bottom panels of Figures~\ref{fig:bpt_wfm} and \ref{fig:bpt_nfm}, were derived from the pPXF emission-line flux errors (rescaled by $\sqrt{\chi^2_\nu}$ where $\chi^2_\nu > 1$) and propagated to the logarithmic line ratios assuming uncorrelated errors, i.e. $\sigma_{\log(F_1/F_2)} = (\ln 10)^{-1}\sqrt{(\sigma_{F_1}/F_1)^2 + (\sigma_{F_2}/F_2)^2}$. H$\alpha$ contours (percentiles 50\% to 90\%) obtained from the MUSE WFM data are overlaid on both WFM and NFM spatial maps as a reference.

In the WFM data (Figure~\ref{fig:bpt_wfm}), the [\ion{N}{II}]-BPT map reveals a centrally concentrated Seyfert nucleus surrounded by a broad ring of composite and star-forming emission at intermediate radii ($\sim 1$--$3\arcsec$), transitioning to composite and then LINER/Seyfert spaxels in the outer region beyond $\sim 3$--$4\arcsec$. The [\ion{S}{II}] and  [\ion{O}{I}] BPT maps classify the vast majority of the field as star-forming, with a compact Seyfert nucleus, and a LINER outer region. The H$\alpha$ contours show that the star-forming and composite ring coincides with the H$\alpha$ ring identified in Figure~\ref{fig:muse_flux_maps}, while the Seyfert nucleus lies within the innermost contour.

The NFM data (Figure~\ref{fig:bpt_nfm}) resolve the inner $\sim 3\farcs7$ ($\sim 1.8$~kpc) at higher spatial resolution ($\mathrm{FWHM}_{\rm PSF} = 0\farcs19$). In the [\ion{N}{II}]-BPT map, the innermost bins are classified as Seyfert, transitioning to composite at $\sim 0.5$--$1\arcsec$ and then to star-forming emission in the outer regions of the NFM field. The [\ion{S}{II}] and  [\ion{O}{I}] BPT maps again classify most bins as star-forming, with a Seyfert core confined to the central $\sim 0\farcs5 - 1\arcsec$. The WFM H$\alpha$ contours overlaid on the NFM maps show that the transition from Seyfert and composite to star-forming emission coincides with the inner edge of the H$\alpha$ ring.

Taken together, the BPT classification across both WFM and NFM fields reveals a three-zone radial structure: a nuclear region whose line ratios require a hard ionising continuum, an intermediate ring of composite and star-forming spaxels at projected distances of $\sim 1$--$3\arcsec$ ($\sim 0.5$--$1.4$~kpc) indicating ongoing star formation, and an outer region at $r \gtrsim 3$--$4\arcsec$ where LINER-like line ratios indicate that the ionisation can no longer be explained by young stars alone, likely reflecting either photoionisation by central SMBH accretion emission at large distances (i.e.\ a light echo from a past accretion episode) or emission from evolved post-AGB stellar populations. Although shock excitation by stellar winds or AGN-driven outflows is often invoked in LINER-like regions \citep{Dopita95}, the remarkably low ionised gas velocity dispersion measured across the field argues against shocks as the dominant ionisation mechanism in this case. The presence of a star-forming ring at intermediate radii is in agreement with the stellar population analysis presented below (see  Section~\ref{sec:results_stellarpops}), where young light-weighted ages are found at comparable distances. 

Differentiation between photoionisation in retired galaxies (galaxies that stopped forming stars) and Seyfert or LINER activity is usually done through the WHAN diagnostic \citep{CidFernandes11}. We note, however, that we did not employ this diagnostic in our analysis. The WHAN diagram is based on the $\mathrm{EW}(\mathrm{H}\alpha)$ and the [\ion{N}{II}] $\lambda$ 6583/H$\alpha$ ratio, which can be affected by the specific conditions of this system. In particular, the [\ion{N}{II}]/H$\alpha$ ratio has a well-known dependence on the gas-phase nitrogen abundance, owing to secondary nitrogen production \citep{Kewley2002, Stasinska2006}. The near-solar stellar metallicities found in the central region of the host Sect.~\ref{sec:results_stellarpops}) suggest that the gas-phase nitrogen abundance may be correspondingly elevated, which would artificially boost the [\ion{N}{II}]/H$\alpha$ ratio and shift spaxels towards the AGN side of the WHAN diagram, independently of the true ionisation mechanism. Given this, we consider the standard BPT diagnostics, which use ratios of lines spanning a wider range of ionisation potentials, and the independent JWST MIRI/MRS analysis presented in Section \ref{jwst_results}, to be more robust classifiers for this source.

\subsubsection{Ionising luminosity and light-echo analysis}
\label{sec:results_lion}

\begin{figure*}[htbp]
\begin{center}
    \includegraphics[width=\linewidth]{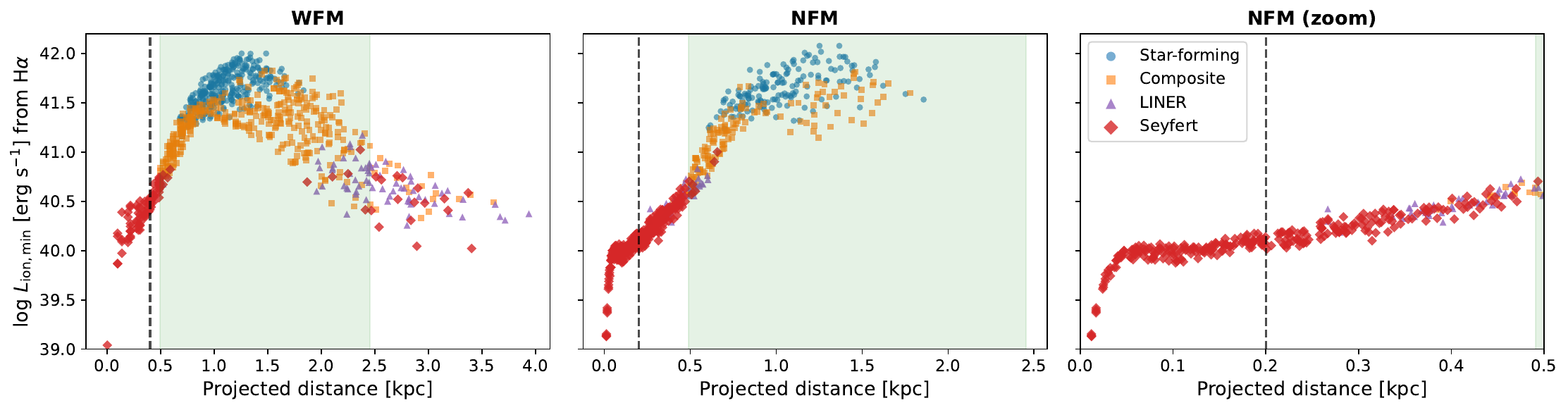}

\caption{Minimum ionising luminosity $L_{\rm ion,min}$ as a function of projected distance, derived from H$\alpha$, for the WFM (left), NFM (centre), and a zoom of the NFM inner 0.5 kpc (right). Points are colour-coded by [\ion{N}{II}]-BPT classification as indicated in the legend. The vertical black dashed line indicates the maximum projected distance at which emission from gas responding to the \textit{Ansky} event is expected to be detected. The green area indicates the region mostly dominated by star-forming or composite bins, which should not be considered to interpret $L_{\rm ion,min}$.
\label{fig:muse_lion}}
\end{center}

\end{figure*}

Following the recombination-balance methodology introduced by \citet{Keel12,Keel17}, extended to a spatially resolved, spaxel-based form by \citet{French23}, and later applied by \citet{Wevers24b} and \citet{Wevers24c} to TDE and QPE host galaxies, we constrain the past ionising output of the nucleus of SDSS1335+0728 by computing the minimum ionising luminosity $L_{\rm ion,min}$ required to power the observed Balmer-line emission in each spatial bin. For each bin, the observed H$\alpha$ or H$\beta$ line luminosity is converted into a rate of hydrogen-ionising photons $Q(\mathrm{H})$ using Case~B recombination coefficients \citep{Osterbrock2006}, and the minimum ionising luminosity is then obtained as \citep[following][]{French23, Wevers24b, Wevers24c}:
\begin{equation}
  L_{\rm ion,min} = \frac{n_r\,L_H\,E_{\rm ion}}{E_H}\,\frac{1}{f},
  \label{eq:lion}
\end{equation}
where $L_H$ is the observed H$\alpha$ or H$\beta$ narrow line luminosity, $E_H$ is the corresponding Balmer-line photon energy, and $E_{\rm ion} = 13.6$~eV is the hydrogen ionisation threshold energy. The factor $n_r = \alpha_B/\alpha_{H,\rm eff}$ is the number of recombinations, or equivalently the number of hydrogen-ionising photons required, per emitted Balmer-line photon under Case~B \citep{Osterbrock2006}. $f$ is the covering fraction of the spaxel as seen from the nucleus, $ f = \frac{\left(2\arctan(0.5/r)\right)^2}{4\pi}$, with $r$ the projected distance from the nucleus in units of spaxels. The resulting $L_{\rm ion,min}$, computed as a function of the projected distance from the nucleus in kpc, is a lower limit of the ionising luminosity primarily because $E_{\rm ion}$ in Equation~\ref{eq:lion} accounts only for hydrogen-ionising photons in the 13.6--54.6~eV range, assuming that higher-energy photons are absorbed by helium \citep{French23, Wevers24b, Wevers24c}; the true bolometric ionising luminosity will be higher if the spectral energy distribution extends significantly above 54.6~eV. Additional factors include the use of projected distances (which provide only a lower limit on the true distance from the nucleus), no intrinsic host galaxy extinction correction on individual spaxels (which is low, see Appendix \ref{app:red_map}), and the implicit assumption of spherical symmetry in Equation~\ref{eq:lion}, whereas the emission-line morphology appears axisymmetric, due to the orientation of the galaxy.

This method assumes that all ionising photons originate from a central nuclear source (whether a classical AGN or another form of SMBH accretion such as a TDE), so that the gas is responding to ionising radiation emitted by the nucleus at an earlier time (a photoionisation light echo). A direct consequence of this assumption is that, in regions where the BPT classification indicates that the dominant ionisation mechanism is local star formation rather than a nuclear source, the inferred $L_{\rm ion,min}$ will be overestimated: the method attributes to the nucleus photons that are predominantly produced locally by young stars, and the covering-fraction correction applied at large projected distances further amplifies this excess. Additionally, the $L_{\rm ion,min}$ estimates at small projected distances from the nucleus should be treated with caution due to a geometric effect in the covering-fraction correction. Specifically, equation~\ref{eq:lion} approximates the spaxel solid angle as the product of two angular widths, which is only valid in the small-angle regime ($r \gg 0.5$~spaxels). As $r \to 0$, each width $2\arctan(0.5/r) \to \pi$ and the expression formally tends to $f \to \pi/4$; the product approximation then ceases to be a faithful solid angle, but the covering fraction is in any case no longer small. In this regime $L_{\rm ion,min}$ (Equation~\ref{eq:lion}) is no longer amplified by the $1/f$ factor and instead converges to $\sim n_r L_H E_{\rm ion}/E_H$, underestimating the true nuclear luminosity required to illuminate gas over all solid angles. The innermost bins should therefore be interpreted with care.

In this light-echo analysis we do not attempt to convert projected distances into light-travel delays. The projected distance does not constrain the delay between the emission of the ionising photons and their reprocessing by the gas, which depends on the unknown position of the gas along the line of sight. Specifically, gas whose reprocessed emission reaches the observer with a given delay relative to the direct nuclear emission lies on an isodelay surface, a paraboloid with the nucleus at its focus, which in projection spans a wide range of radii \citep[see][for a detailed discussion]{Mummery25}. Each radial bin therefore combines emission from gas illuminated at different epochs. However, the isodelay geometry places a robust upper limit on the projected distance that radiation emitted since the onset of the \textit{Ansky} event can have reached. For radiation emitted a time $\Delta t$ before the observations, and gas extending out to a radius $R_{\rm cloud}$, the maximum projected distance is $X_{\rm max} \approx (2R_{\rm cloud}\,c\,\Delta t)^{1/2}$ (equation~28 in \citealt{Mummery25}). Adopting $c\,\Delta t \approx 5.5$~light-years (December 2019 to mid-2025) and $R_{\rm cloud} \approx 4$~kpc (a conservative estimate of the galaxy semi-major axis, considering that the Sloan Digital Sky Survey -- SDSS; \citealt{York00} -- Petrosian radius at 90\% in the r-band is 7\farcs08 or $\sim3.4$~kpc), we obtain $X_{\rm max} \approx 0.12$~kpc. This value is conservative, as it neglects the recombination time and requires gas located almost exactly along the line of sight in front of the nucleus. Emission from gas responding to the \textit{Ansky} event is further spread by the PSF, with $\text{PSF}_{\text{NFM}}=0\farcs19$ ($\approx 0.09$~kpc) and $\text{PSF}_{\text{WFM}}=0\farcs62$ ($\approx 0.29$~kpc). Adopting $X_{\rm max}$ plus one PSF FWHM as the maximum projected distance at which emission from gas responding to the \textit{Ansky} event can be detected, accretion-driven ionisation detected beyond $\sim$0.2~kpc ($\sim$0\farcs4) in the NFM data and beyond $\sim$0.4~kpc ($\sim$0\farcs85) in the WFM data can therefore not be attributed to the \textit{Ansky} event, and must trace nuclear activity prior to December 2019.

Figure~\ref{fig:muse_lion} presents $L_{\rm ion,min}$ as a function of projected distance for the H$\alpha$-estimate, combining the WFM (left column), NFM (centre column), and a zoomed view of the NFM inner $\sim$0.5~kpc (right column). Each spatial bin is colour-coded by its [\ion{N}{II}]-BPT classification. The vertical black dashed line in each panel marks the maximum projected distance at which emission from gas responding to the \textit{Ansky} event is expected to be detected, including PSF smearing (bins beyond this line trace nuclear activity prior to December 2019). In Appendix \ref{app:lion_hbeta} we show the results for the H$\beta$-estimate, which are in agreement with the results derived from H$\alpha$.

The ionising luminosity profile shows a clear radial trend, rising steeply from $\log L_{\rm ion,min} \sim 40$ at the smallest resolved scales to a maximum of $\sim 42$ at projected distances of $\sim$1--1.5~kpc, and declining to $\sim$40.5 at the largest projected distances sampled by the WFM ($r \sim 3$--$4$~kpc). However, as noted above, the bins classified as star-forming or composite in the BPT diagrams (zone highlighted in green in Figure \ref{fig:muse_lion}) are not tracing nuclear ionisation, and their $L_{\rm ion,min}$ values should not be interpreted as constraints on the nuclear luminosity. In addition, bins very close to the centre are affected by the geometric effects described above, and bins within the dashed line in Figure~\ref{fig:muse_lion} may include emission from gas responding to the \textit{Ansky} event itself; these bins are therefore excluded from the following discussion.

Restricting attention to the Seyfert- and LINER-classified bins (outside the green area in Figure \ref{fig:muse_lion}) located beyond the region that can have been illuminated by the \textit{Ansky} event, the inferred $L_{\rm ion,min}$ spans $\log L_{\rm ion,min} \sim 40$--$40.7$~erg~s$^{-1}$ between projected distances of $\sim$0.2 and 0.5~kpc (NFM), while the outermost bins ($r \gtrsim 2.5$~kpc, WFM) remain consistent with $\log L_{\rm ion,min} \sim 40.5$~erg~s$^{-1}$. We stress that $L_{\rm ion,min}$ is, by definition, a lower limit on the ionising luminosity needed to reproduce the observed line emission. Since the Seyfert-classified bins extend to $\sim$0.5~kpc from the nucleus and are also found at $\sim$2.5--3.5~kpc, and LINER-classified bins extend to $\sim$4~kpc, beyond the region that radiation from the \textit{Ansky} event can have reached (including PSF smearing), these values indicate that the nucleus reached an ionising luminosity of at least $\log L_{\rm ion,min} \sim 40.5$~erg~s$^{-1}$ prior to December 2019. However, owing to the isodelay-surface geometry, this analysis cannot establish for how long this precursor activity was sustained, nor whether the true luminosity increased, declined, or remained constant over time, unless we make some assumptions about the 3D distribution of the excited gas, which go beyond the scope of this work.

Independent constraints on the present-day nuclear luminosity are available from multiwavelength observations. The current monochromatic luminosity at 5100\,\AA, derived by integrating the nuclear power-law continuum flux within an aperture of one NFM PSF, is $\log \lambda L_{\lambda}(5100) = 41.3$~erg~s$^{-1}$\, while the quiescent X-ray luminosity at $0.3$--$2$~keV from contemporaneous XMM-Newton observations is $\log L_{\rm X} = 40.46$~erg~s$^{-1}$ \citep{Chakraborty26A}. Since the ZTF light curve (Figure~\ref{fig:ztf_lc}) shows that the \textit{Ansky} event temporarily elevated the nuclear flux and that the source is now declining towards its pre-2019 quiescent state, these July 2025 measurements can be considered upper limits on the quiescent X-ray (excluding QPE flares) and optical monochromatic luminosity prior to 2019. Comparing these values directly with $L_{\rm ion,min}$ requires caution, as the latter is a strict lower limit and converting $L_{\rm X}$ or $L_{5100}$ into bolometric luminosities involves assumptions that may not hold for \textit{Ansky}. A more informative comparison comes from the SED modelling of \citet{Sanchez-Saez24}, which showed that a big-blue-bump (BBB) component with $\log L_{\rm BBB} = 42.21_{-4.37}^{+0.16}$~erg~s$^{-1}$ was required to explain the GALEX ultraviolet photometry obtained $\sim$15 years before the \textit{Ansky} event (however note the large uncertainties due to the limited UV observations); this pre-existing accretion-disc luminosity is consistent within its (large) errors with the $L_{\rm ion,min}$ values measured in the Seyfert-classified bins.

In conclusion, the results presented in Figure~\ref{fig:muse_lion}, combined with the SED analysis of \citet{Sanchez-Saez24}, point to accretion-driven ionising flux, prior to 2019, although its duration cannot be constrained with the present analysis. The implications of these results are further discussed in Section \ref{discussion}.

\begin{figure*}[htb!]
\begin{center}
    \includegraphics[width=\linewidth]{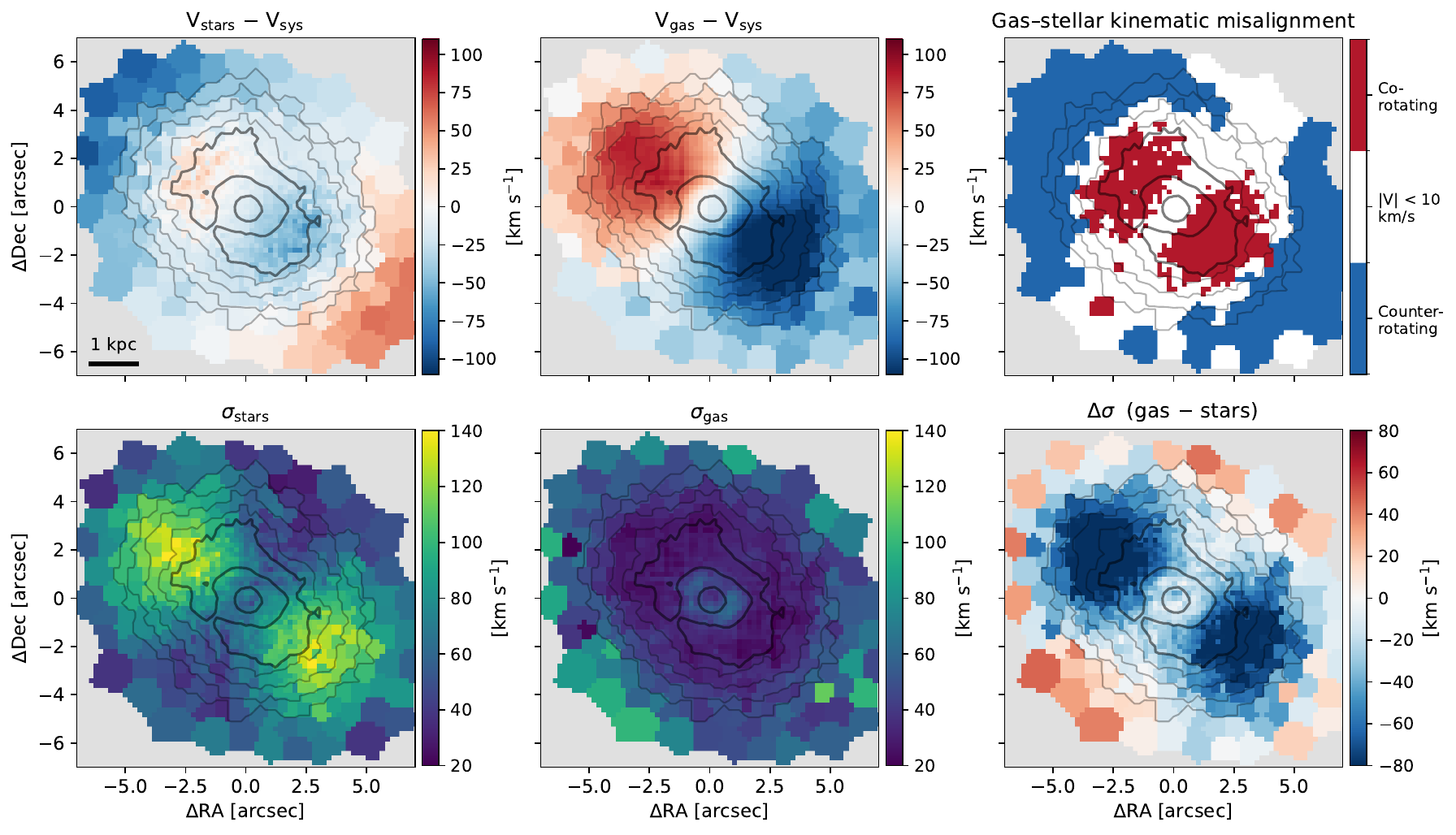}

\caption{Spatially resolved kinematic maps from the MUSE WFM datacube. Top row: stellar velocity $V_{\rm stars} - V_{\rm sys}$ (left), gas velocity $V_{\rm gas} - V_{\rm sys}$ (centre), and gas--stellar kinematic misalignment map (right), where blue and red indicate counter- and co-rotating spaxels respectively, and white marks spaxels with $|V| < 10$~km~s$^{-1}$. Bottom row: stellar velocity dispersion $\sigma_{\rm stars}$ (left), gas velocity dispersion $\sigma_{\rm gas}$ (centre), and their difference $\Delta\sigma = \sigma_{\rm gas} - \sigma_{\rm stars}$ (right). All panels show the H$_{\alpha}$ emission line contours (percentiles 50\% to 90\%). The brightest region is highlighted with a darker colour.  
\label{fig:muse_kinematics}}
\end{center}
\end{figure*}

\subsubsection{Stellar and gas kinematics}
\label{sec:results_kinematics}

Figure~\ref{fig:muse_kinematics} shows the spatially resolved kinematic maps derived from the \texttt{pPXF} fits to the WFM datacube. The upper row presents the stellar line-of-sight velocity field $V_{\rm stars} - V_{\rm sys}$ (left), the ionised gas velocity field $V_{\rm gas} - V_{\rm sys}$ (centre), and a kinematic misalignment map (right) indicating the sign agreement between stellar and gas velocities. The lower row shows the stellar velocity dispersion $\sigma_{\rm stars}$ (left), the gas velocity dispersion $\sigma_{\rm gas}$ (centre), and the difference $\Delta\sigma = \sigma_{\rm gas} - \sigma_{\rm stars}$ (right). All panels show the H$_{\alpha}$ emission line contours for reference.

The stellar velocity field reveals a striking inversion of the rotation sense with radius: the inner $\sim 2\arcsec$ rotates in one direction, whilst the outer region beyond $\sim 3$--$4\arcsec$ rotates in the opposite sense, with a projected amplitude of $\sim
100$~km~s$^{-1}$ on either side of the kinematic axis. The ionised gas, in contrast, shows a smooth, coherent rotation pattern across the entire field without such an inversion. This difference is clearly captured in the gas--stellar kinematic misalignment map (Figure~\ref{fig:muse_kinematics}, top right), where the outskirts of the galaxy are classified as counter-rotating, indicating that the gas and the outer stellar region rotate in opposite senses. Notably, the peak of the H$_{\alpha}$ emission coincides with the region of co-rotation between the stars and the gas. A similar configuration has been observed in the counter-rotating nuclear ring of other galaxies, like NGC~7742, where a young stellar population co-rotates with the gas in the ring while the old stellar body counter-rotates \citep{Martinsson18}.

The stellar velocity dispersion map shows a pronounced peak of $\sigma_{\rm stars} \sim 130$~km~s$^{-1}$ at the radius where the velocity inversion occurs, coinciding spatially with the transition between the two counter-rotating stellar components; this enhancement is consistent with the superposition of two stellar discs rotating in opposite directions along the line of sight, whose opposing motions broaden the observed velocity distribution (the so-called double $\sigma$ or $2\sigma$ feature, with $\sigma$ referring to the stellar velocity dispersion; \citealt{Krajnovic2011}). The gas velocity dispersion is notably low across the entire field ($\sigma_{\rm gas} \lesssim 60$~km~s$^{-1}$), resulting in a strongly negative $\Delta\sigma$ in the central region; such kinematically cold gas has also been observed in other TDE and QPE host galaxies \citep[e.g.][]{Wevers24b, Wevers24c}.

The combination of a counter-rotating stellar component and the low gas velocity dispersion ($\sigma_{\rm gas} \lesssim 60$~km~s$^{-1}$) indicates that the gas is dynamically cold and settled, consistent with the aftermath of a minor merger as proposed for other QPE hosts \citep{Wevers24c}. A low $\sigma_{\rm gas}$ alone does not discriminate between a minor merger, smooth misaligned cold accretion, or the tidal stripping of a gas-rich companion. However, the undisturbed morphology in the continuum (white image in Figure \ref{fig:muse_flux_maps}) strongly disfavours a major merger, while the counter-rotation requires an external source of angular momentum misaligned with the pre-existing stellar body, favouring a wet minor merger in which accreted gas settled into a counter-rotating region \citep[e.g.][]{Eliche-Moral11, Corsini2014, Pizzella18,Martinsson18}.

Finally, we estimated the SMBH mass of SDSS1335+0728 using the $M-\sigma$ relation of \cite{McConnell2013} and isolating the region within 1\farcs5 from the nucleus (to avoid the counter-rotating area). For the WFM, we obtained a stellar velocity dispersion of $\sigma_{\text{stars}}=76$ km s$^{-1}$, which implies a SMBH mass of $8.93\times10^5 M_{\odot}$ (with a $1\sigma$ range of $3.00\times10^5 M_{\odot}$ and $2.66\times10^6 M_{\odot}$). We also estimated the SMBH mass using the NFM-derived stellar velocity dispersion within the same  1\farcs5 radius, obtaining $\sigma_{\text{stars}}=75.4$ km s$^{-1}$, which implies a SMBH mass of $8.5\times10^5 M_{\odot}$ (with a $1\sigma$ range of $2.85\times10^5 M_{\odot}$ and $2.54\times10^6 M_{\odot}$). Both results are consistent with the estimation presented in \cite{Sanchez-Saez24} of $\sim 10^6 M_{\odot}$, and with the mass range of all known QPE hosts \citep{Arcodia21,Wevers24b}.

\begin{figure*}[htb]
    \includegraphics[width=\linewidth]{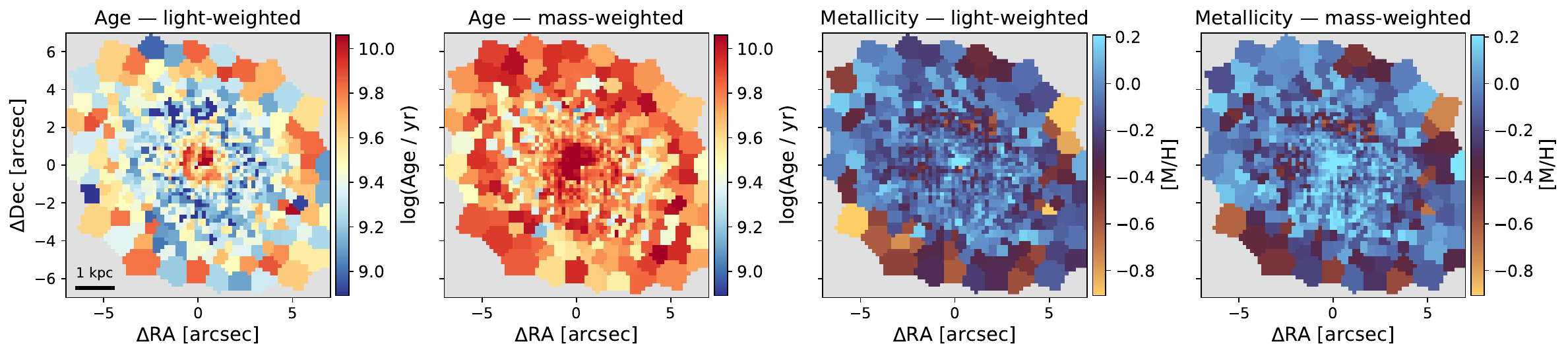}

 \caption{Spatially resolved stellar population maps from the MUSE WFM datacube, derived from the \texttt{pPXF} template weights. From left to right: light-weighted mean age, mass-weighted mean age (both in $\log_{10}(\mathrm{Age}/\mathrm{yr})$), light-weighted mean metallicity, and mass-weighted mean metallicity ($[\mathrm{M/H}]$).
\label{fig:muse_stellar_pops}}
\end{figure*}

\subsubsection{Stellar population properties}
\label{sec:results_stellarpops}

Figure~\ref{fig:muse_stellar_pops} shows the spatially resolved stellar population maps derived from the \texttt{pPXF} template weights using the MILES SSP library: light-weighted mean age (left), mass-weighted mean age (centre-left), light-weighted mean metallicity (centre-right), and mass-weighted mean metallicity (right). We note that the galaxy shows negligible reddening overall (see section \ref{app:red_map} of the Appendix).

The light-weighted age map reveals an intermediate-age stellar population across much of the field, with $\log_{10}(\mathrm{Age}_L/\mathrm{yr}) \sim 9$--$10$, and the youngest ages concentrated in the ring classified as star-forming in the BPT diagrams (Section \ref{sec:results_bpt}). The mass-weighted age map is systematically older ($\log_{10}(\mathrm{Age}_M/\mathrm{yr}) \gtrsim 9.5$--$10.0$), indicating that the host galaxy is dominated by an old stellar population typical of a quiescent low-redshift galaxy. 

The metallicity maps reveal a clear radial gradient in both the light- and mass-weighted estimates: the central $\sim 2$--$3\arcsec$ shows near-solar metallicity ($[\mathrm{M/H}] \sim -0.1$ to $0.0$), declining to sub-solar values ($[\mathrm{M/H}] \sim -0.4$ to $-0.6$) in the outer regions. 

The presence of a relatively young light-weighted population in the star-forming ring superimposed on an old mass-weighted host is broadly consistent with the minor-merger scenario discussed in Section~\ref{sec:results_kinematics}, in which accreted gas may have triggered a localised episode of star formation. This is consistent with hydrodynamic simulations of gas-rich minor mergers, which show that stripped gas can settle into long-lived rings and sustain star formation over several Gyr, thereby rejuvenating the host stellar population \citep{Mapelli15}.

Together with the BPT classification presented in Section~\ref{sec:results_bpt}, these stellar population properties confirm that SDSS1335+0728 still hosts ongoing star formation, albeit confined to the ring structure rather than the nucleus.

\subsection{JWST MIRI/MRS data cube analysis}\label{jwst_results}

\subsubsection{Nuclear mid-infrared spectrum}
\label{sec:jwst_1D}

\begin{figure*}[htbp]
\begin{center}
    \includegraphics[width=\linewidth]{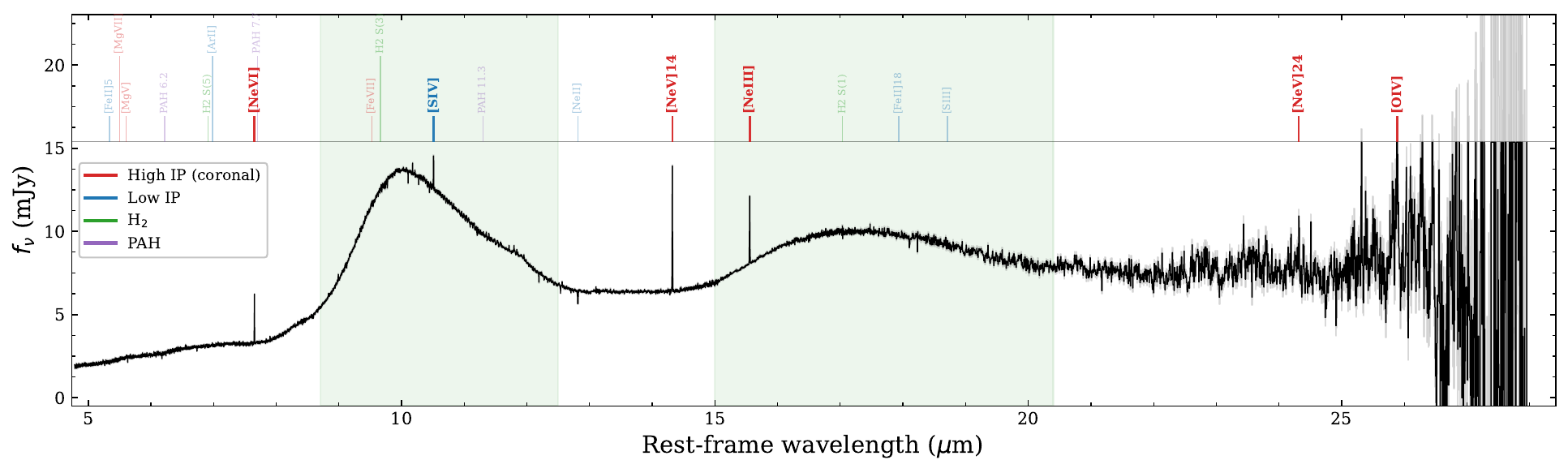}

    \caption{JWST MIRI/MRS rest-frame spectrum of SDSS\,J1335$+$0728. The spectrum is dominated by broad silicate emission peaking near 10\,$\mu$m and 17--18\,$\mu$m (green regions). Narrow emission lines are marked by vertical ticks above the spectrum, colour-coded by line category: high-IP coronal lines (red), low-IP ISM and star-formation tracers (blue), \Htwo\ rotational lines (green), and PAH features (purple). Detected lines (SNR\,$\geq 3$) are shown with bold, fully opaque labels; non-detections are shown at reduced opacity. The grey shaded band indicates the $1\sigma$ flux uncertainty. The bumpy structure red-ward of 22\,$\mu$m is likely fringing/detector noise.}
    \label{fig:jwst_nucler_spec}

\end{center}
\end{figure*}

The integrated 1D MIRI/MRS spectrum of SDSS\,J1335$+$0728 is shown in Figure~\ref{fig:jwst_nucler_spec}. The most striking feature is a broad emission excess peaking near 10\,$\mu$m, which we identify as the 9.7\,$\mu$m silicate emission feature. The continuum rises steeply from $\sim$2\,mJy at 5\,$\mu$m to $\sim$14\,mJy at the silicate peak, then declines through a local minimum near 13\,$\mu$m before rising again to a second, shallower broad excess around 17--20\,$\mu$m, consistent with the 18\,$\mu$m silicate feature also seen in emission.

We measure the silicate feature strength using two independent continuum estimation methods, following \cite{Goold26}, as $S_{\rm sil} = \ln(f_{\rm obs}/f_{\rm cont})$ at 9.7\,$\mu$m ($S_{\rm sil}$[9.7$\mu$m]) by fitting a power-law continuum. Additionally, following \cite{Masterson25}, we fitted the continuum with a cubic spline, and measured $S_{\rm sil}$[9.7$\mu$m] and $S_{\rm sil}$ at 18$\mu$m ($S_{\rm sil}$[18$\mu$m]). Further details are provided in Section \ref{app:silicate} of the Appendix. Both methods yield consistent results, with $S_{\rm sil}$[9.7$\mu$m]~$=0.99$ for the \cite{Goold26} definition and with  $S_{\rm sil}$[9.7$\mu$m]~$=1.12$ and  $S_{\rm sil}$[18$\mu$m]~$=0.36$ for the \cite{Masterson25}.

Silicate emission is typically associated with dust that is directly illuminated by a central source and reemits thermally in the MIR, viewed through a low effective MIR optical depth. This condition can be met by two distinct dust configurations: a dust distribution that is globally optically thin in the MIR, or a clumpy distribution of individually optically thick clouds \citep{Nenkova08}. Previous MIR observations of AGN show a broad range of silicate strengths, with the features seen in emission in both type~1 and type~2 sources \citep[e.g.][]{Hatziminaoglou15,Garcia-Bernete24,Ramos-Almeida25,Ramos-Almeida26}, and clumpy torus models can reproduce silicate emission of comparable strength to that measured in the nuclear MIR spectrum of \textit{Ansky} \citep[see e.g. Fig.~10 of][]{Hatziminaoglou15}. Since host-galaxy contamination tends to weaken the silicate emission \citep{Hatziminaoglou15}, our measurement is unlikely to overestimate the nuclear silicate strength. The measured $S_{\rm sil}$[9.7$\mu$m]~$\approx1.0$ thus lies at the upper end of the AGN distribution (typically $S_{9.7}\lesssim0.5$ in emission; \citealt{Hatziminaoglou15,Ramos-Almeida25}).

Given \textit{Ansky}'s host potential previous weak state (see Sect.~\ref{sec:results_lion}), we draw parallels to MIR observations of LLAGN (persistent dust illumination picture) and TDE hosts (where dust is newly heated). In the LLAGN regime, the classical torus structure may be absent due to insufficient radiative support from disc-driven winds \citep{Elitzur2006}. In this type of galaxies, \citet{Goold26} detected broad, spatially unresolved silicate emission at $\sim$9.7\,$\mu$m in five of the ReveaLLAGN targets, attributing it to optically thin dust localised on parsec scales near the central engine, and consistent with remnants of a dissipating torus or diffuse nuclear dust \citep{Mason2013}. For the case of TDEs, \citet{Masterson25} presented the first JWST MIRI/MRS observations of TDE candidates, finding that all four MIR-selected TDEs in their sample exhibit silicate emission features at $\sim$9.7 and 18\,$\mu$m that are significantly stronger than those of all AGN in their comparison samples. Clumpy torus models can reproduce silicate features of comparable strength, but these TDE candidates lack detectable optical flares, implying a line of sight that is optically thick at optical wavelengths. This is difficult to reconcile with the relatively unobscured view of directly illuminated clouds that such strong silicate emission requires in clumpy models. \citet{Masterson25}, therefore, attribute the strong silicate emission to the absence of an optically thick, clumpy torus around dormant SMBHs, and show that the MIR continuum is well reproduced by a time-dependent, optically thin dust model. This argument cannot be applied to \textit{Ansky}: the detection of the optical/UV transient \citep{Sanchez-Saez24}, the featureless HST UV continuum \citep{Zhu25}, and the soft X-ray emission \citep{Hernandez-Garcia25a,Chakraborty25b} all indicate a relatively unobscured line of sight towards the nucleus. This is consistent both with a globally optically thin dust distribution and with a clumpy distribution of optically thick clouds viewed through low-covering regions. The silicate features alone therefore do not constrain the dust geometry in \textit{Ansky}, and a clumpy, torus-like dust component cannot be excluded. The silicate emission in SDSS\,J1335$+$0728 is thus consistent with both the LLAGN and the TDE pictures, which could also be simultaneously correct in the case of a TDE occurring in a pre-existing LLAGN.

In addition, superimposed on the silicate continuum, Figure \ref{fig:jwst_nucler_spec} shows the detection of several narrow emission lines spanning a wide range of ionisation potentials (IPs). The most prominent detections are the high-IP lines [\ion{Ne}{VI}]\,$\lambda$7.65\,$\mu$m (IP~$= 126$\,eV), [\ion{Ne}{V}]\,$\lambda$14.32\,$\mu$m (IP~$= 97$\,eV), [\ion{Ne}{III}]\,$\lambda$15.56\,$\mu$m (IP~$= 41$\,eV), and [\ion{O}{IV}]\,$\lambda$25.89\,$\mu$m (IP~$= 55$\,eV), as well as the low-IP line [\ion{S}{IV}]\,$\lambda$10.51\,$\mu$m (IP~$= 35$\,eV), which can be produced by both accretion-driven activity and star formation. The detection of lines with IPs well above the $\sim$55\,eV limit of hot stellar populations (in particular [\ion{Ne}{VI}] at 126\,eV and [\ion{Ne}{V}] at 97\,eV) provides strong evidence for hard photoionisation by an accretion-powered nuclear source. Whether this hard ionising field is dominated by the \textit{Ansky} event or by the pre-existing accretion activity is discussed in Sect.~\ref{ansky_discussion}.  

The PAH features at 6.2 and 7.7\,$\mu$m are weak or undetected (opaque labels in Figure \ref{fig:jwst_nucler_spec}) in the nuclear spectrum of Figure \ref{fig:jwst_nucler_spec}. This may be expected in accretion-dominated nuclei, where the hard radiation field can process or destroy the small dust grains that carry the PAH emission \citep[e.g.][]{Sajina22}, but may also result from intrinsically weak star formation within the extraction aperture, which lies interior to the star-forming ring identified with MUSE (Sect.~\ref{sec:results_bpt}). The weak PAH features are therefore consistent with, but do not by themselves demonstrate, a hard, accretion-dominated nuclear radiation field. A detailed analysis of the PAH emission across the host galaxy will be presented in a forthcoming paper.

\subsubsection{Spatially resolved emission-line maps}
\label{sec:jwst_maps}

\begin{figure*}[htbp]
\begin{center}
        \includegraphics[width=\linewidth]{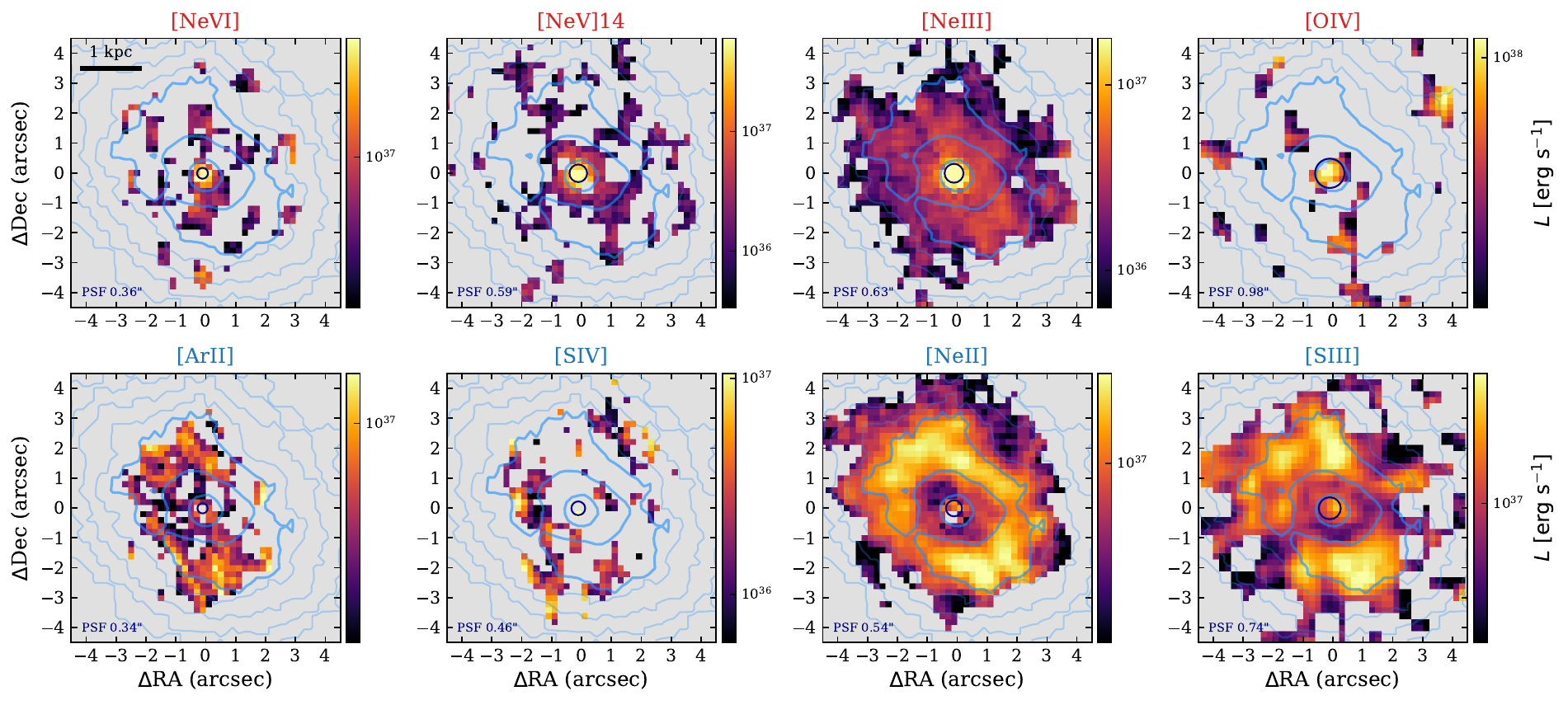}

    \caption{Spatially resolved line luminosity maps for eight emission lines detected in the MIRI/MRS spectral cubes of SDSS\,J1335$+$0728, displayed on a logarithmic colour scale. The top row shows high-IP lines, normally associated with accretion-driven activity: ([\ion{Ne}{VI}]$\lambda$7.65, [\ion{Ne}{V}]$\lambda$14.32, [\ion{Ne}{III}]$\lambda$15.56, and [\ion{O}{IV}]\,$\lambda$25.89\,$\mu$m). The bottom row shows low-IP tracers ([\ion{Ar}{II}]$\lambda$6.99, [\ion{S}{IV}]$\lambda$10.51, [\ion{Ne}{II}]$\lambda$12.81, [\ion{S}{III}]$\lambda$18.71). All maps are reprojected onto a common 0.20\,arcsec\,pixel$^{-1}$ grid and shown over a $9\times9$\,arcsec FoV centred on the nucleus. Spaxels below SNR\,$= 3$ are masked. The PSF FWHM at the relevant observed wavelength is indicated by the circle (centred on the MIRI/MRS nucleus, defined as the [\ion{Ne}{III}] peak) and in the lower-left corner of each panel. Blue contours show the MUSE WFM H$\alpha$ emission (percentiles 50\% to 90\%), aligned to the MIRI/MRS nucleus.}
    \label{fig:jwst_maps}
\end{center}

\end{figure*}

Figure~\ref{fig:jwst_maps} shows the spaxel-by-spaxel line luminosity maps for eight emission lines detected at SNR\,$\geq 3$ in the MIRI/MRS spectral cubes. Other lines are also detected  (including  H$_2$ S(1), which mimics the behavior observed for [\ion{Ne}{II}]$\lambda$12.81), but we leave their analysis for future work. All maps are displayed on a common $9\arcsec \times 9$\arcsec~FoV centred on the nucleus, with the PSF FWHM at the relevant observed wavelength indicated in each panel. We note that the PSF FWHM varies by a factor of $\sim$3 across the MIRI/MRS wavelength range (from $\approx$0\farcs34 at [\ion{Ar}{II}] to $\approx$0\farcs98 at [\ion{O}{IV}]). To compare the MIR and optical structures directly, Figure~\ref{fig:jwst_maps} also shows the MUSE WFM H$\alpha$ contours, aligned to the MIRI/MRS nucleus. Since the MIRI/MRS FoV increases with wavelength (Sect.~\ref{jwst_data}), the mapped lines sample different fractions of the MUSE structures: all of them cover the nucleus, the channel~2--4 lines cover most or all of the star-forming ring, and only the channel~4 lines ([\ion{S}{III}], [\ion{O}{IV}]) reach the innermost part of the outer LINER-like region. None of the MIRI/MRS maps sample the outer region beyond $\sim$2~kpc, so all constraints on it rely on MUSE alone.

The maps reveal a clear dichotomy between the spatial distributions of high-IP and low-IP lines. The low-IP tracers, [\ion{Ne}{II}]\,$\lambda$12.81\,$\mu$m and [\ion{S}{III}]\,$\lambda$18.71\,$\mu$m, are spatially extended across the host galaxy, with emission detected out to $\sim$4\arcsec ($\sim$2 kpc) from the nucleus, and concentrated in a ring-like structure that coincides with the H$\alpha$ ring traced by the MUSE contours. This emission is consistent with ongoing star formation in a ring, in agreement with the optical MUSE diagnostics (section~\ref{sec:results_bpt}). [\ion{Ar}{II}]\,$\lambda$6.99\,$\mu$m and [\ion{S}{IV}]$\lambda$10.51 appear to trace similar extended structure, although they are generally much weaker. [\ion{Ne}{III}]\,$\lambda$15.56\,$\mu$m, which traces gas at intermediate IP (41\,eV), shows a hybrid morphology: a luminous nuclear concentration and detectable extended emission at radii of 1--3\arcsec, suggesting contributions from both SMBH accretion photoionisation in the nucleus and young H\,\textsc{ii} regions.

In contrast, the high-IP coronal lines [\ion{Ne}{VI}]\,$\lambda$7.65\,$\mu$m, [\ion{Ne}{V}]$\lambda$14.32, and [\ion{O}{IV}]\,$\lambda$25.89\,$\mu$m are predominantly concentrated in the central resolution element and show only sparse, low-significance detections at larger radii. These outer detections may arise from the PSF wings, noise, genuinely extended emission, or a combination of these effects; the morphology of these lines is therefore consistent with an unresolved nuclear component. [\ion{Ne}{V}] and [\ion{Ne}{VI}] require a hard, accretion-powered ionising source, whereas [\ion{O}{IV}] may also have contributions from shocks or very hard stellar populations.

\subsubsection{[\ion{Ne}{III}]/[\ion{Ne}{II}] ionisation diagnostic}
\label{sec:ne3ne2}

\begin{figure}[htbp]
\begin{center}
        \includegraphics[width=0.85\linewidth]{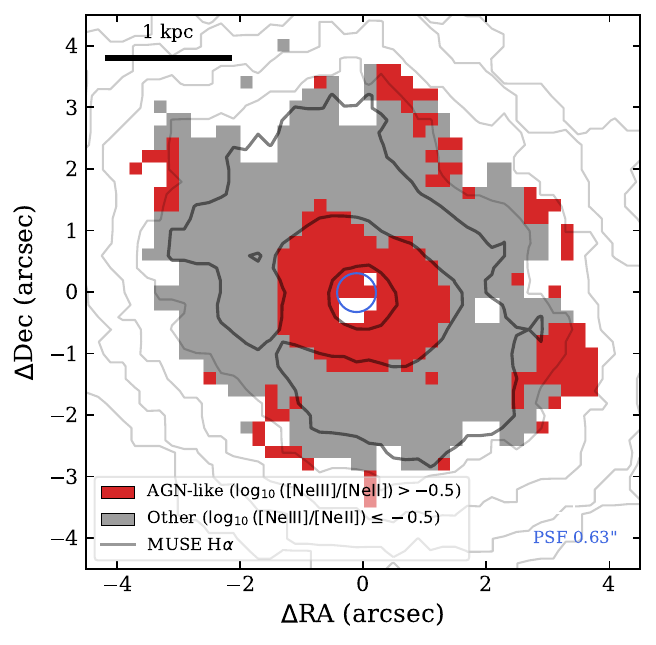}

    \caption{$\log_{10}([\ion{Ne}{III}]/[\ion{Ne}{II}])$ classification map of SDSS\,J1335$+$0728. Spaxels with $\log_{10}([\ion{Ne}{III}]/[\ion{Ne}{II}]) > -0.5$ (red) are classified as AGN-like (accretion-driven), whilst spaxels at or below this threshold (grey) are consistent with star-formation-dominated ionisation. Only spaxels with SNR\,$\geq 3$ in both lines are shown. The blue circle indicates the PSF FWHM of [\ion{Ne}{III}], centred on the MIRI/MRS nucleus. Black contours show the MUSE WFM H$\alpha$ emission (percentiles 50\% to 90\%).}
    \label{fig:jwst_ne_class}
\end{center}
\end{figure}

The $\log_{10}([\ion{Ne}{III}]/[\ion{Ne}{II}])$ line-ratio map is shown in Figure~\ref{fig:jwst_ne_class}. This ratio is a widely used MIR ionisation diagnostic that is largely insensitive to dust extinction and traces the hardness of the local radiation field \citep[e.g.][]{Inami2013,Feuillet2025}. We adopt a threshold of $\log_{10}([\ion{Ne}{III}]/[\ion{Ne}{II}]) = -0.5$ to separate AGN-dominated from star-formation-dominated spaxels, following \cite{Feuillet2025}. The ratio is computed only in spaxels where both lines independently satisfy the peak SNR\,$\geq 3$ criterion (Appendix~\ref{app:jwst_linefitting}); spaxels failing this criterion in either line are left blank. We note that this line-ratio diagnostic employs two transitions falling within the same MRS channel (channel~3), for which the PSF FWHM differs by only $\sim$15\% (0.54 vs.\ 0.63\,arcsec), minimising any systematic bias from resolution mismatch \citep[see][for a discussion]{Zhang2024}. The two maps were reprojected onto a common pixel grid but not convolved to a common PSF, so the slightly broader [\ion{Ne}{III}] PSF may redistribute nuclear [\ion{Ne}{III}] emission to larger radii more efficiently than [\ion{Ne}{II}], potentially biasing the ratio upwards. In addition, the SNR\,$\geq 3$ requirement implies a statistical uncertainty on the logarithmic ratio of up to $\approx$0.2~dex, so spaxels lying close to the threshold cannot be individually classified with confidence.

The map shows a well-defined central region of AGN-like (accretion-driven) ratios (red in Figure~\ref{fig:jwst_ne_class}), extending $\sim$1\farcs5 ($\sim$0.7 kpc) from the nucleus. Beyond this radius, the line ratio drops below the threshold, and the emission is consistent with ionisation by stellar populations, while some of the spaxels of the most external regions show tentative evidence of accretion-driven ionisation. This spatial stratification is fully consistent with the line luminosity maps: the high-IP line emission is predominantly concentrated in the central region where the Neon ratio indicates accretion-driven conditions, while the extended [\ion{Ne}{II}] and [\ion{S}{III}] emission traces the star-forming host beyond the AGN-photoionised zone. The H$\alpha$ contours in Figure~\ref{fig:jwst_ne_class} show that the AGN-like region lies interior to the star-forming ring, whose position coincides with the star-formation-dominated (grey) spaxels, while the few AGN-like spaxels at the edge of the field lie beyond the H$\alpha$ ring, at the inner edge of the LINER-like outer region identified with MUSE. The MIR diagnostics, therefore, provide independent support for the optical BPT-based result: the accretion-driven regions are spatially restricted to the areas interior to the star-forming ring and beyond it.

\section{Discussion}\label{discussion}

In the following sections, we discuss the implications of the results presented in Section \ref{results} for the accretion history of SDSS1335+0728 and the origin of the \textit{Ansky} event, as well as for the QPE population in general.

\subsection{Implications for the origin of the \textit{Ansky} event}\label{ansky_discussion}

The light-echo analysis (Section~\ref{sec:results_lion}) constrains the pre-2019 ionising output of the nucleus. Restricting attention to the accretion-driven-classified bins, where the assumption of a central ionising source is physically justified, a minimum ionising luminosity of $\log L_{\rm ion,min} \approx 40.5$~erg~s$^{-1}$ is required to explain the Balmer-line emission at projected distances well beyond the region that radiation from the \textit{Ansky} event can have reached ($\gtrsim$0.2~kpc in the NFM data, and up to $\sim$4~kpc in the WFM data). Two caveats temper further interpretation. First, $L_{\rm ion,min}$ is a strict \emph{lower} limit on the nuclear ionising luminosity. Second, the projected distance does not constrain the light-travel delay, and the isodelay-surface geometry smears the correspondence between projected distance and epoch \citep{Mummery25}, so the duration of the precursor activity cannot be constrained. The derived $\log L_{\rm ion,min}$ therefore implies that a dramatically brighter past is \emph{not required} to explain the extended emission: the data are fully accounted for by a nuclear source with an ionising luminosity of $\gtrsim 10^{40.5}$~erg~s$^{-1}$ active prior to December 2019. This luminosity lies $\sim 2$~dex below the pre-2019 BBB luminosity from the SED modelling of \citet{Sanchez-Saez24}, and above the BBB lower limit (2.5\% limit) reported in the same work ($\log L_{\rm bol} = 37.84$~erg~s$^{-1}$). The natural reading is that the nucleus was not dormant prior to December 2019, and that the \textit{Ansky} event temporarily elevated the luminosity above a pre-existing baseline. The ZTF light curve (Figure~\ref{fig:ztf_lc}) independently shows the source declining back towards, but not yet reaching, its pre-2019 flux level.

The emission-line maps (Sections~\ref{sec:muse_fluxmaps} and~\ref{sec:jwst_maps}) reveal a clear spatial structure. The low-IP tracers ([\ion{Ne}{II}], [\ion{S}{III}], [\ion{Ar}{II}]) and [\ion{O}{III}] are spatially extended across the host, confirming widespread ionisation well beyond the nucleus, whereas the high-IP lines ([\ion{Ne}{VI}], [\ion{Ne}{V}], [\ion{O}{IV}]) are predominantly concentrated in the central resolution element, consistent with an unresolved nuclear component. This differential compactness is expected for high-IP lines, which require a high ionisation parameter close to the source, independently of the age of the ionising source (e.g., \citealt{Mummery25}). High-IP neon lines are expected to arise from dense gas ($n_{\rm H} \sim 10^{4}$--$10^{7}$~cm$^{-3}$) within $\sim$10--100 light-years of the ionising source \citep[Section~7 and Figure~21 of][]{Mummery25}, where recombination times are short \citep{Osterbrock2006, Mummery25}, so this emission responds rapidly to changes in the ionising continuum. Moreover, the region that radiation emitted since December 2019 can have reached ($X_{\rm max} \approx 0.12$~kpc; Section~\ref{sec:results_lion}) is comparable to or smaller than the MIRI/MRS resolution element at the relevant wavelengths ($\mathrm{FWHM}_{\rm PSF} \approx 0\farcs36$--$0\farcs98$, i.e.\ $\approx$0.2--0.5~kpc). However, as the expected extent of the coronal-line-emitting gas lies well below this resolution, and such emission can remain bright for decades to centuries after illumination \citep{Mummery25}, our data do not distinguish whether the coronal-line emission is powered by the \textit{Ansky} event, by the pre-existing nuclear activity, or by both. In contrast, the low- and intermediate-IP lines are detected out to projected distances of $\sim$2--3~kpc, without showing any preferential direction, well beyond the region that the \textit{Ansky} event can have illuminated, requiring a source of ionising photons active prior to December 2019.

The strong 9.7 and 18\,$\mu$m silicate features observed in emission (Section~\ref{sec:jwst_1D}) indicate dust directly illuminated by the central source and viewed through a low effective MIR optical depth. The inferred silicate strengths lie closer to the MIR-selected TDE population of \citet{Masterson25} than to most AGN \citep{Hatziminaoglou15,Ramos-Almeida25}, although they do not by themselves exclude a clumpy, torus-like dust component \citep{Nenkova08, Ramos-Almeida26}. Moreover, the suppression or absence of a thick torus is itself predicted at these low luminosities ($L_{\rm bol}\lesssim10^{42}$\,erg\,s$^{-1}$; \citealt{Elitzur2006}), so the dust properties alone do not discriminate between a TDE and an LLAGN origin for the precursor activity.

We therefore consider two scenarios for the pre-2019 accretion history. In both pictures, the \textit{Ansky} event corresponds to a slow and faint transient event associated with a $\sim 10^6 M_{\odot}$ SMBH with previously ongoing accretion, which has implications for the evolution observed (e.g., interaction between the accreted and precursor material; \citealt{Zhang25}). We note however that observationally, the available data do not decisively favour any single scenario:
\begin{itemize}

    \item \textit{Scenario A: an LLAGN}. A nucleus accreting at a low Eddington ratio prior to 2019 reproduces the $L_{\rm ion,min}$ profile and the pre-2019 BBB emission required by the GALEX photometry \citep{Sanchez-Saez24}, while the silicate emission is compatible with the dissipating or absent torus expected at these luminosities \citep{Elitzur2006}. Crucially, the weak, compact [\ion{Ne}{V}] is consistent with JWST observations of LLAGNs \citep{Goold26}. For $L_{\rm bol} \approx 10^{40.5}$~erg~s$^{-1}$ and $M_{\rm BH} \approx 10^6\,M_\odot$, the Eddington ratio is $\lambda_{\rm Edd} \approx 2\times10^{-4}$ ($\log\lambda_{\rm Edd} \approx -3.6$), at or below the threshold of $\log(L_{\rm bol}/L_{\rm Edd}) = -3.5^{+0.6}_{-1.0}$ below which the ReveaLLAGN sample of \citet{Goold26} falls systematically below the $L_{\rm [\ion{Ne}{V}]}$--$L_{\rm X}$ relation. We note that since $L_{\rm ion,min}$ is a lower limit on the ionising luminosity, and the bolometric luminosity will in general exceed the ionising one, this Eddington ratio is also a lower limit.

    \item \textit{Scenario B: a long-lived TDE remnant disc.} The framework of \citet{Mummery25} demonstrates that TDE discs can power EELRs, coronal lines, and MIR line ratios that mimic a classical AGN on standard diagnostic diagrams, without requiring one. A freely evolving remnant declines as $L_{\rm bol} \propto t^{-1.2}$ \citep{Cannizzo90, Mummery25}, while the disc temperature cools more slowly ($T_p \propto t^{-0.3}$), sustaining ionising radiation for $t_{\rm ION} \sim 200$~yr. Light-travel-time effects (isodelay contours) and the long recombination timescales of low-density ($n_{\rm H} \sim 10^{1}$--$10^{3}$~cm$^{-3}$) gas then allow a single disruption to produce observable EELR features for $10^{3}$--$10^{4}$~yr after the TDE disc itself has faded. The energetics are consistent: a single TDE disc radiates $\sim10^{52}$--$10^{53}$~erg in ionising photons over $t_{\rm ION}\sim$ few hundred years \citep{Mummery25}, well above what is required to sustain $L_{\rm ion,min}\approx10^{40.5}$~erg~s$^{-1}$. However, as the duration of the precursor activity is unconstrained, our analysis cannot distinguish a single earlier disruption from repeated and/or partial tidal disruptions.

\end{itemize}

Recent works have claimed either a connection between faded AGNs and TDE and QPE events, or a connection between turn-on AGNs and TDEs. Our results can be associated with both. The gas-rich minor-merger host, the $M_{\rm BH}\sim10^6\,M_{\odot}$, and the kpc-scale EELR match the properties that other works have associated with recently faded AGNs in QPE and TDE hosts \citep[e.g.,][]{Jiang25,Xiong25}. Scenario~A, however, does not require a faded nucleus: the $L_{\rm ion,min}$ profile is consistent with low-level accretion prior to the \textit{Ansky} event, without requiring a previously luminous phase. This places \textit{Ansky} at an earlier stage than the fully faded hosts, but possibly within the same evolutionary picture. More speculatively, both scenarios A and B could be linked to the ``starfall'' mechanism of \citet{McKernan22}, in which the loss of orbital angular momentum during the first $\sim 0.1$~Myr of an AGN phase drives a heavy rain of stars onto the SMBH. In Scenario~A, such a process could feed an ongoing low-level nucleus, while in Scenario~B it could supply the disrupted star. However, this mechanism is not directly tested by our data, and other pathways connecting external gas accretion to nuclear activity are equally viable. The kinematically misaligned gas observed in SDSS1335+0728 points to an environment favourable for gas inflow and SMBH fuelling \citep[see][for a review]{Storchi-Bergmann19}, but does not by itself establish a causal link with the \textit{Ansky} event.

\subsection{Implications for the QPE population}\label{qpe_discussion}

IFS studies of QPE hosts have shown EELRs and elevated [\ion{O}{III}] emission \citep{Wevers24c,Xiong25}. \cite{Jiang25} attributed these to recently faded AGNs. SDSS1335+0728 complicates this picture. As shown above, the fossil ionisation tracers in \textit{Ansky}'s host are degenerate between an ongoing LLAGN (Scenario~A) and a long-lived TDE remnant disc (Scenario~B), and neither viable scenario requires a previously luminous, now-totally-faded nucleus. At the population level, this implies that the EELR and elevated [\ion{O}{III}] used to infer faded AGN are not, on their own, a unique signature of fading: an ongoing low-level or transient accretion source reproduces the same fossil signatures. We refrain from concluding that the population-level faded-AGN interpretation must be revised on the basis of a single object. Spatially resolved analyses of further QPE hosts with VLT/MUSE NFM or JWST MIRI/MRS are needed before the population-level scenario can be reassessed.

The lack of QPEs detected before February 2024 also deserves comment. It is worth recalling how poorly the earlier epochs were sampled. The QPEs in \textit{Ansky} are days-long soft X-ray flares that recur roughly every $4-15$ days, with varying period and flaring time scales, yet the pre-2019 soft X-ray record of SDSS1335+0728 amounts to only a handful of archival snapshots, with no sensitive dedicated monitoring, so a non-detection over this period tells us very little about whether QPEs were actually occurring. If the absence is nonetheless real, the disc--orbiter framework \citep[e.g.][]{Linial23a,Franchini23,Lu23} offers a natural way to account for it. The simplest possibility ties directly to Scenario~A of Section~\ref{ansky_discussion}: if the pre-2019 flow was a geometrically thick RIAF at $\lambda_{\rm Edd} \approx 10^{-4}$, collisions between the orbiter and the disc would have produced much weaker and softer flares than the present-day eruptions, most likely below any detectable level. The 2019 event would then have assembled the thin, dense disc that the QPE mechanism requires, which would explain why the eruptions only appeared after the transient. Other configurations could produce the same outcome: the earlier disc may simply have been too compact to be struck at the orbiter's pericentre, or the orbiter may have shared the plane of the pre-existing disc, so that no detectable eruptions arose until the 2019 event built a new, misaligned disc.

The minor-merger or external gas accretion history inferred from the host kinematics offers a plausible dynamical context for these channels, as perturbations of the nuclear stellar dynamics could enhance the delivery of stars to the SMBH, providing both TDEs and potential QPE orbiters. However, the present data do not constrain the timing of the interaction relative to the nuclear transients. The star-forming ring cannot by itself date the interaction, since gas accreted in minor mergers can sustain ring star formation over several Gyr \citep{Mapelli15}. Establishing a causal connection would require quantitative constraints on the interaction epoch and on the relevant dynamical transport timescales. We therefore regard this connection as a plausible evolutionary scenario rather than an observational result.

It is also worth considering whether the QPEs themselves might have contributed significantly to the observed circumnuclear ionisation. The QPEs in \textit{Ansky} are estimated to emit $\gtrsim 10^{48}$~erg per burst, and constraints from time-resolved spectroscopy of the X-ray outflow estimate a mass budget of $\sim 10^{-3} M_\odot$/burst for a kinetic energy $\sim 10^{48.5-49}$ erg/burst \citep{Chakraborty25b,Chakraborty26b}. Dividing the radiated energy per burst by an average recurrence time of $\sim 10$ days yields a time-averaged radiated luminosity of $\sim 10^{42}$ erg s$^{-1}$, emitted predominantly in the soft X-rays, which well exceeds the $L_{\rm ion,min}$ estimates of Figure~\ref{fig:muse_lion}. However, given the recent turn-on in QPE activity ($\lesssim 5$ yr), the ongoing QPE phase could not have contributed to the thousands light-year scales probed by the MUSE and MIRI/MRS data, although it may have partially contributed to the emission of unresolved gas close to the nucleus, including the nuclear coronal lines. On the other hand, if \textit{Ansky} exhibited prior QPE phases (akin to e.g. GSN 069; \citealt{Miniutti23b}), it is possible that the radiation from those prior active periods contributed. However, such a long QPE lifetime appears unlikely in \textit{Ansky}, given the large mass budget inferred from the outflows \citep{Chakraborty25b,Chakraborty26b}, which place an upper limit of $\lesssim 1000$ eruptions before total energy budget of the orbiter-SMBH system is depleted (assuming the currently observed mass loss per burst, energetics, and recurrence properties). We thus consider it unlikely that the QPEs themselves contribute significantly to the observed ionisation profile.

\section{Summary}\label{conclusions}

We have presented spatially resolved optical (VLT/MUSE WFM and NFM) and mid-infrared (JWST MIRI/MRS) spectroscopy of SDSS1335+0728, the host galaxy of the \textit{Ansky} event and the most extreme known QPE source. Our main findings are as follows.

The stellar kinematics reveal a counter-rotating outer region and kinematically decoupled gas with extremely low velocity dispersion ($\sigma_{\rm gas} \lesssim 60$~km~s$^{-1}$), consistent with a past minor merger involving retrograde gas accretion (with respect to the original rotation sense of the galaxy). The accreted gas fuelled the ring-like star formation confirmed independently by the BPT diagnostics (Section~\ref{sec:results_bpt}), by the younger light-weighted stellar ages at comparable radius (Section~\ref{sec:results_stellarpops}), and by the spatially extended low-IP MIR emission (Section~\ref{sec:jwst_maps}). The near-solar central metallicity superimposed on an old, metal-poorer outer region is likewise consistent with merger-triggered star formation. This minor-merger history places SDSS1335+0728 among the gas-rich, post-merger environments that are preferentially found among TDE and QPE hosts \citep{Wevers24b, Wevers24c}, although its ongoing star formation distinguishes it from the post-starburst and quiescent Balmer-strong galaxies that are over-represented among these hosts relative to the general galaxy population. The $M$--$\sigma$ relation yields $M_{\rm BH} \sim 10^6\,M_\odot$, in agreement with previous estimates and with the mass range of known QPE hosts.

The optical BPT diagnostic maps identify a three-zone radial structure: a nucleus ionised by SMBH accretion, an intermediate star-forming ring at $\sim$0.5--1.4~kpc confirmed by the youngest light-weighted stellar ages and extended low-IP MIR emission, and a LINER-like outer region. The MIR [\ion{Ne}{III}]/[\ion{Ne}{II}] map independently supports the accretion-driven nucleus and the star-forming ring, and tentatively the accretion-driven ionisation at the inner edge of the outer region. High-IP lines ([\ion{Ne}{VI}], [\ion{Ne}{V}], [\ion{O}{IV}]) are predominantly concentrated in the nuclear resolution element, consistent with an unresolved nuclear component. [\ion{Ne}{V}] and [\ion{Ne}{VI}] demonstrate a hard, accretion-powered ionising field, although the present data do not distinguish whether it is dominated by the \textit{Ansky} event or by the pre-existing accretion activity.

A Balmer-line light-echo analysis yields a minimum ionising luminosity of $\log L_{\rm ion,min} \sim 40.5$~erg~s$^{-1}$ across the accretion-driven-classified bins, at projected distances well beyond the region that radiation from the \textit{Ansky} event can have reached ($\gtrsim$0.2~kpc, and up to $\sim$4~kpc). This demonstrates that the nucleus was not dormant before December 2019, although the duration of this precursor activity cannot be constrained. Rather, the \textit{Ansky} event temporarily elevated the luminosity above a pre-existing baseline. This is consistent with the SED analysis presented in \citet{Sanchez-Saez24}, in which BBB emission was required to account for the UV emission prior to 2019.

Strong silicate emission at 9.7 and 18 $\mu$m indicates directly illuminated dust viewed through a low effective MIR optical depth, arising either from globally optically thin dust or from a clumpy distribution of optically thick clouds. Its 9.7 $\mu$m strength lies at the upper end of the AGN distribution, closer to that of MIR-selected TDEs, and it is compatible with both TDEs and LLAGNs with dissipating tori.

We propose two viable scenarios for the pre-2019 accretion history: (A) an LLAGN with sustained low-level accretion ($\log\lambda_{\rm Edd} \gtrsim -3.6$), not yet dormant, or (B) a long-lived TDE remnant disc powering the extended emission. In both scenarios, the \textit{Ansky} event corresponded to a slow and faint transient in a $\sim 10^6\,M_{\odot}$ SMBH with already ongoing accretion. Both offer an alternative to the ``faded AGN'' interpretation recently invoked for QPE host galaxies: if Scenario~B applies more broadly, the faded-AGN signatures proposed for other QPE hosts may instead trace prior stellar disruptions. The minor-merger history of SDSS1335+0728 identifies a host environment favourable for gas inflow and SMBH fuelling, providing a plausible context for its nuclear transient activity.

\section*{Data availability}
The VLT/MUSE deep stacks can be obtained at the ESO Science Portal via \url{https://archive.eso.org/dataset/ADP.2025-11-17T13:54:49.781}. The JWST MIRI/MRS data can be obtained at the MAST archive via \url{https://mast.stsci.edu/search/ui/#/jwst} searching for programme ID 8245.

\begin{acknowledgements}

We thank the referee for a careful reading of the manuscript and comments that led to its
improvement. 
The authors acknowledge the use of Claude Opus 4.6 and Opus 5.5 (Anthropic) as a language editing tool and for code debugging during the preparation of this manuscript.

We acknowledge support from ANID-Chile BASAL CATA FB210003 (FEB, RJA) and Millennium Science Initiative Program NCN$2023\_002$ (PA, JCu, PL, MLMA). 
We acknowledge support from ANID-Chile FONDECYT Regular 1241005 (FEB, PA), 1241422 (PA, PL, FEB), 1251444 (JCu), 1231418 (FAV), 1231718 (RJA), and FONDECYT Iniciaci\'on 11241477 (LHG). 
RA is funded by the Gordon and Betty Moore Foundation (Grant \#13526) and John Templeton Foundation (Grant \#63445).  The opinions expressed in this publication are those of the author(s) and do not necessarily reflect the views of these Foundations.
SB acknowledges the National Agency for Research and Development (ANID) grant Gemini-32240014.
YD acknowledges the ANID / Fondo GEMINI 2025 / 32250023.
PA acknowledges the CAV, CIDI N. 21 U. de Valparaíso, Chile.
MG is funded by Spanish MICIU/AEI/10.13039/501100011033 and ERDF/EU grant PID2023-147338NB-C21.
GM acknowledges support from grant PID2023-147338NB-C21 funded by MICIU/AEI/10.13039/501100011033 and ERDF/EU.
LHG, GM, MG acknowledge support from ESA through the Science Faculty - Funding reference ESA-SCI-E-LE-383.
MS acknowledges the Czech Science Foundation (GA\v{C}R) grant no. 26-23342I. 
D.I. and A.B.K acknowledge funding provided by the University of Belgrade— Faculty of Mathematics (contract 451-03-33/2026-03/200104) through grants by the Ministry of Education, Science, and Technological Development of the Republic of Serbia.

This work is based on observations made with the NASA/ESA/CSA James Webb Space Telescope. The data were obtained from the Mikulski Archive for Space Telescopes at the Space Telescope Science Institute, which is operated by the Association of Universities for Research in Astronomy, Inc., under NASA contract NAS 5-03127 for JWST. These observations are associated with program \#8245.

Based on observations collected at the European Southern Observatory under ESO programmes 115.28E2.001 and 115.28E2.002.

Based on observations obtained with the Samuel Oschin 48-inch Telescope at the Palomar Observatory as part of the \textit{Zwicky} Transient Facility project. ZTF is supported by the National Science Foundation under Grant No. AST-1440341 and a collaboration including Caltech, IPAC, the Weizmann Institute for Science, the Oskar Klein Center at Stockholm University, the University of Maryland, the University of Washington, Deutsches Elektronen-Synchrotron and Humboldt University, Los Alamos National Laboratories, the TANGO Consortium of Taiwan, the University of Wisconsin at Milwaukee, and Lawrence Berkeley National Laboratories. Operations are conducted by COO, IPAC, and UW.

Based on observations obtained with the Samuel Oschin Telescope 48-inch and the 60-inch Telescope at the Palomar Observatory as part of the \textit{Zwicky} Transient Facility project. ZTF is supported by the National Science Foundation under Grants No. AST-1440341 and AST-2034437 and a collaboration including current partners Caltech, IPAC, the Weizmann Institute for Science, the Oskar Klein Center at Stockholm University, the University of Maryland, Deutsches Elektronen-Synchrotron and Humboldt University, the TANGO Consortium of Taiwan, the University of Wisconsin at Milwaukee, Trinity College Dublin, Lawrence Livermore National Laboratories, IN2P3, University of Warwick, Ruhr University Bochum, Northwestern University and former partners the University of Washington, Los Alamos National Laboratories, and Lawrence Berkeley National Laboratories. Operations are conducted by COO, IPAC, and UW.

\end{acknowledgements}

\bibliographystyle{aa}
\bibliography{bibliography}

\begin{appendix}
\nolinenumbers

\section{MUSE Stellar Continuum and Emission-Line Fitting with \texttt{pPXF}}
\label{app:ppxf}

Each MUSE WFM and NFM Voronoi bin was fitted using \texttt{pPXF} \citep{Cappellari2023}. Stellar kinematics and emission-line fluxes were extracted simultaneously fitting stellar continuum templates drawn from the MILES SSP library \citep{Vazdekis2010} with a \citet{Kroupa2001} IMF and BaSTI isochrones \citep{Pietrinferni2004}. The fits were done over the rest-frame range $4600$--$7200$~\AA. Prior to fitting, the MILES templates (native resolution $\mathrm{FWHM}_{\mathrm{MILES}} = 2.51$~\AA) were convolved to the wavelength-dependent MUSE LSF following \citet{Bacon2017}, with the observed-frame LSF FWHM de-redshifted to the rest-frame before computing the quadrature difference; spectra and templates were then log-rebinned onto a common velocity scale with a template-to-galaxy oversampling ratio of~2. The fit included a degree-4 additive Legendre polynomial to account for residual continuum shape mismatches, and stellar kinematics were recovered with two kinematic moments ($V$ and $\sigma$; bounds $[-500,+500]$ and $[10,500]$~km~s$^{-1}$). We note that no differences in the fits are found when using four Gauss--Hermite moments ($V$, $\sigma$, $h_3$, $h_4$).

\begin{figure}[hbp]
\begin{center}
\begin{tabular}{c}
    \includegraphics[width=\linewidth]{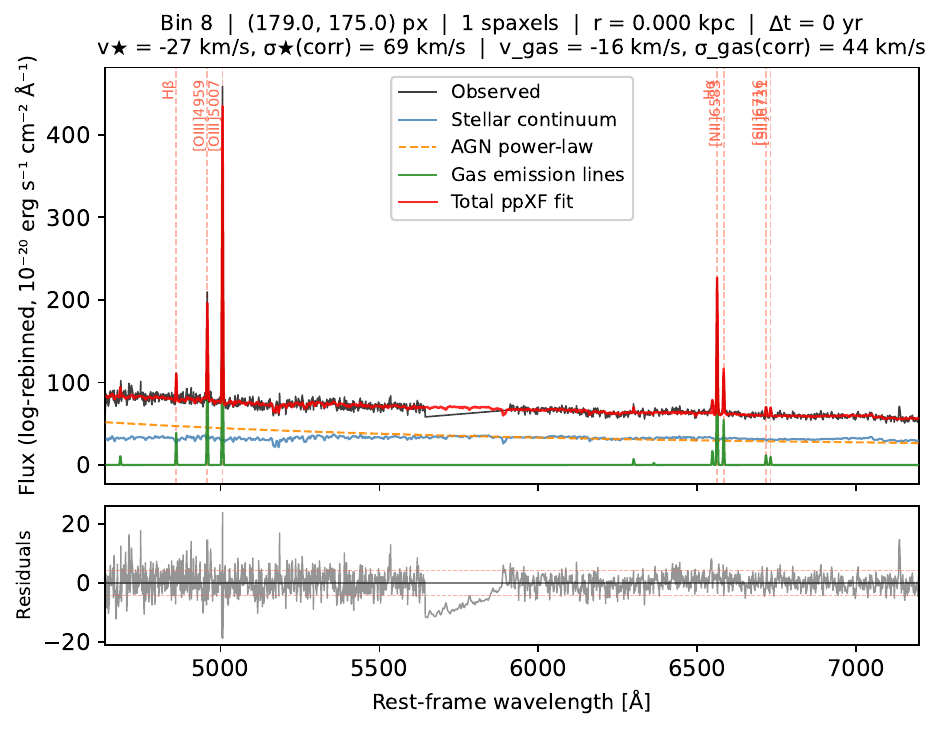}
\end{tabular}
    \caption{Representative \texttt{pPXF} spectral fit for a central NFM bin. The plot shows the observed, log-rebinned spectrum (black), the total \texttt{pPXF} best fit (red), decomposed into the stellar continuum (blue), accretion-driven power-law continuum (orange dashed), and narrow emission-line (green) components. The residuals (observed minus best fit) are shown in the lower sub-panel, with dashed red lines indicating $\pm1\sigma_{\rm rms}$. Vertical dashed lines mark the rest-frame wavelengths of the principal emission lines fitted. The stellar velocity $v_\star$, velocity dispersion $\sigma_\star$, gas velocity $v_{\rm gas}$, and gas velocity dispersion $\sigma_{\rm gas}$ recovered for each bin are indicated in the panel title, together with the bin centre coordinates, number of contributing spaxels, and projected distance from the nucleus.}
    \label{fig:ppxf_fits}
\end{center}
\end{figure}

ZTF photometry contemporaneous with the MUSE observations confirms ongoing accretion activity; thus, an additional accretion-driven continuum component was included in the fit of the central region ($r < 0\farcs5$ for the NFM, and $r < 1\arcsec$ for the WFM), modelled as a linear combination of power laws $F_\lambda \propto \lambda^{\,\alpha}$ with slopes $\alpha \in \{-3.5,\,-3.0,\,-2.5,\,-2.0,\,-1.5,\,-1.0,\,-0.5,\,-0.1\}$ normalised at $5500$~\AA; outside this aperture the power-law component was omitted.

Narrow emission-line templates for H$\beta$, [\ion{O}{III}]$\lambda\lambda$4959,5007, [\ion{O}{I}]$\lambda\lambda$6300,6364, [\ion{N}{II}]$\lambda\lambda$6548,6583, H$\alpha$, and [\ion{S}{II}]$\lambda\lambda$6716,6731 were generated simultaneously via \texttt{emission\_lines} and fitted with two kinematic moments ($V$, $\sigma$; bounds $[-1500,+1500]$ and $[1,600]$~km~s$^{-1}$), with no constraints imposed on flux ratios; bad pixels were linearly interpolated prior to log rebinning and excluded from the fit via the \texttt{goodpixels} mask. As for the stellar component, we note that no differences in the fits are found when using four Gauss--Hermite moments ($V$, $\sigma$, $h_3$, $h_4$).

\begin{figure}[thbp]
\begin{center}
\begin{tabular}{c}
    \includegraphics[width=\linewidth]{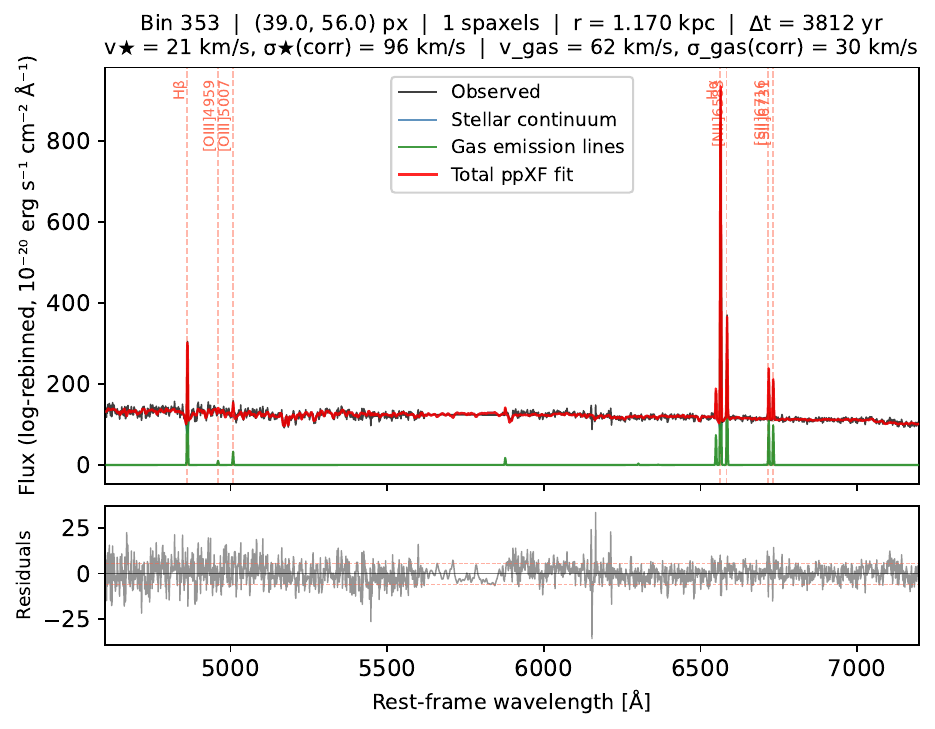}\\
      \includegraphics[width=\linewidth]{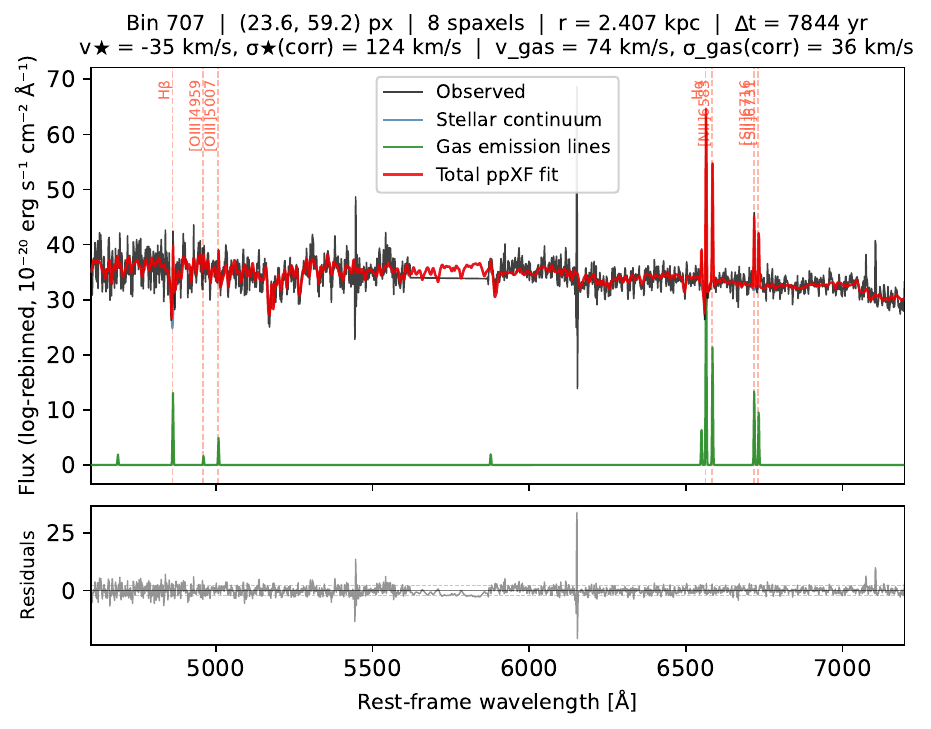}
\end{tabular}
    \caption{As in \ref{fig:ppxf_fits}, but for two WFM bins. No accretion-driven power-law continuum is included as the projected distance from the nucleus is larger than 1\farcs}
    \label{fig:ppxf_fits2}
\end{center}
\end{figure}

The reduced-$\chi^2$ of the WFM \texttt{pPXF} fits per bin has a median, 2.5\% percentile, and 97.5\% percentile of 1.01, 0.75, and 2.54, respectively. For the NFM, the reduced-$\chi^2$ has a median, 2.5\% percentile, and 97.5\% percentile of 4.73, 1.32, and 23.23, respectively, however when restricting the analysis to the inner ring within a radius of 1\farcs5, these correspond to 3.11, 1.24, 7.57, respectively.

Figures \ref{fig:ppxf_fits} and \ref{fig:ppxf_fits2} show representative \texttt{pPXF} spectral fits for a selection of spatial bins spanning the full radial range of the galaxy. The stellar continuum, accretion-driven power-law continuum (nuclear bins only), and narrow emission-line components are shown separately, demonstrating the quality of the simultaneous decomposition across regions of varying ionisation and stellar population properties.

\section{MUSE PSF characterisation using the accretion-driven power-law continuum}
\label{app:psf}

The PSF of the WFM observations was estimated empirically from unsaturated field stars detected within the MUSE FoV using \texttt{DAOStarFinder} \citep{Stetson1987} available in \texttt{photutils}\footnote{\url{https://photutils.readthedocs.io/en/stable/index.html}}, which identifies point sources via a Gaussian-derivative convolution kernel; each detected source was fitted with a two-dimensional Gaussian profile to derive the PSF FWHM. From this, we estimated a PSF of 0\farcs62 at 6720~\AA{} (at the H$_{\alpha}$ region).

\begin{figure}[htbp]
\begin{center}
\includegraphics[width=0.9\linewidth]{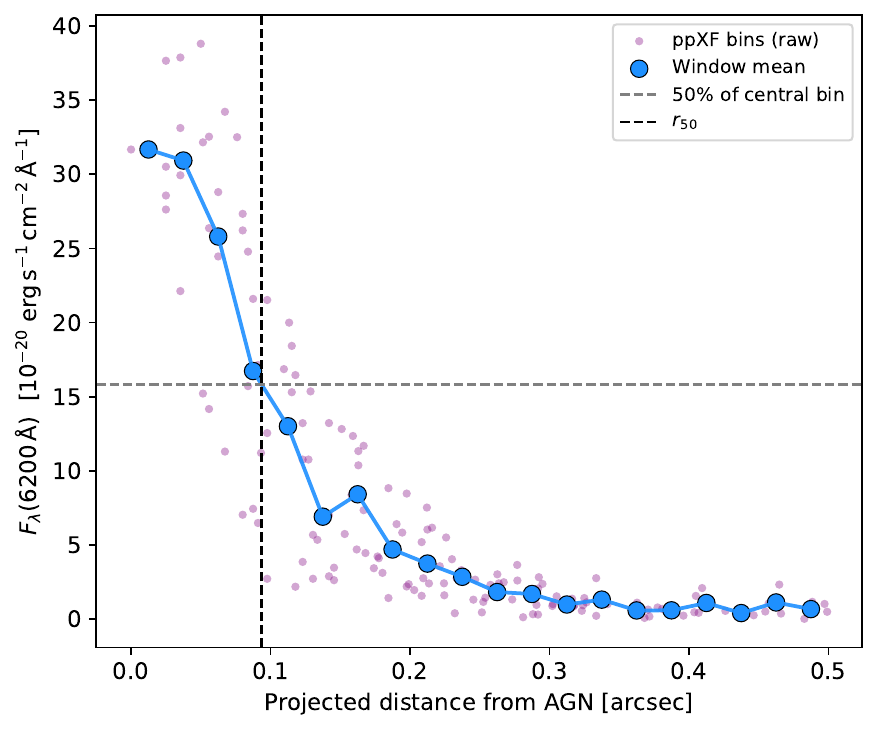}
\caption{Radial profile of the accretion-driven power-law continuum flux density at $\lambda_{\rm ref} = 6200$~\AA{} as a function of projected distance from the nucleus, used to characterise the PSF of the MUSE NFM
observations. Individual \texttt{pPXF} spatial bins are shown as purple circles, and the azimuthally averaged profile in radial windows of $\Delta r = 0\farcs025$ (NFM) is shown as blue circles connected by a solid line. The horizontal dashed grey line marks 50 per cent of the central bin flux density, and the vertical dashed black line indicates the corresponding half-flux radius $r_{50}$, which we adopt as the PSF half-width at half-maximum (HWHM), giving $\mathrm{FWHM}_{\rm PSF} \approx 2\,r_{50} = 0\farcs$19.}
\label{fig:agn_pl}
\end{center}
\end{figure}

As an independent consistency check, we also characterised the PSF from the spatial profile of the AGN power-law continuum component recovered by \texttt{pPXF} in the nuclear bins. Since the AGN is unresolved at the WFM resolution, its power-law continuum flux density at $6200$~\AA{}, evaluated per spatial bin and averaged in radial windows of $\Delta r = 0\farcs2$ (one WFM pixel), traces the PSF directly; the half-flux radius $r_{50}$ of this profile therefore provides an estimate of the PSF half-width at half-maximum (HWHM), and thus $\mathrm{FWHM}_{\mathrm{PSF}} \approx2\,r_{50}$. From this, we obtained a PSF of 0\farcs64 at $6200$~\AA{} (rest-frame). Therefore, both methods yield consistent PSF FWHM estimates, validating the power-law profile approach.

For the NFM observations, no point sources are detected within the $7\farcs5 \times 7\farcs5$ FoV. We therefore adopted $2\,r_{50}$ of the AGN power-law continuum profile as the NFM PSF FWHM estimate, using radial windows of $\Delta r = 0\farcs025$, obtaining a PSF for the NFM of 0\farcs19. Figure \ref{fig:agn_pl} shows the radial profile fitting of the accretion-driven power-law continuum for the NFM.

\section{JWST MIRI/MRS spectral fitting}\label{app:jwst_linefitting}

Emission-line fluxes and kinematic parameters were measured from the stitched 1D spectrum for a set of 17 targeted transitions, comprising high-IP coronal lines ([\ion{Mg}{VII}], [\ion{Mg}{V}], [\ion{Ne}{VI}], [\ion{Fe}{VII}], [\ion{S}{IV}], [\ion{Ne}{V}]$\lambda$14.32, [\ion{Ne}{III}]$\lambda$15.56, [\ion{Ne}{V}]$\lambda$24.32, [\ion{O}{IV}]$\lambda$25.89), low-IP ISM and star-formation tracers ([\ion{Fe}{II}]$\lambda$5.34, [\ion{Ar}{II}]$\lambda$6.99,
[\ion{Ne}{II}]$\lambda$12.81, [\ion{Fe}{II}]$\lambda$17.94, [\ion{S}{III}]$\lambda$18.71), and \Htwo\ rotational lines (S(5)$\lambda$6.91, S(3)$\lambda$9.67, S(1)$\lambda$17.04). Each line was fitted independently with a single Gaussian profile plus a local linear continuum, determined from side-band regions lying between $\pm700$ and $\pm3500$\,\kms\ from the line centre via a weighted least-squares fit using the inverse variance as weights, within a total fitting window of $\pm4000$\,\kms. The continuum-subtracted spectrum was fitted using \texttt{scipy.optimize.curve\_fit} \citep{Virtanen2020}, with the amplitude constrained to be positive and the maximum Gaussian dispersion limited to 8000\,\kms. From the best-fit parameters we derived the integrated line flux, the FWHM, and the velocity offset from systemic. The SNR was defined as the fitted amplitude divided by the rms of the continuum residuals in the side-band windows. A line was considered detected if it simultaneously satisfied SNR\,$\geq 3$, $|\Delta v| \leq 1000$\,\kms, and FWHM\,$\geq 100$\,\kms.

Spatially resolved emission-line maps were produced by applying the same single-Gaussian fitting procedure independently at each spaxel of the MIRI/MRS spectral cubes. The local noise was estimated per spaxel from the rms of the continuum residuals, and spaxels with non-zero data-quality flags in more than half the channels within the line window were excluded. Fits yielding FWHM\,$< 100$\,\kms\ were rejected as unresolved noise features, and only spaxels with peak SNR\,$\geq 3$ were retained. Cube surface brightnesses (MJy\,sr$^{-1}$) were converted to line fluxes per spaxel (erg\,s$^{-1}$\,cm$^{-2}$\,spaxel$^{-1}$) using the spaxel solid angle derived from the \texttt{CDELT1} keyword of each cube header, which varies from 0\farcs13 in channel~1 to 0\farcs35 in channel~4. All maps were reprojected onto a common astrometric grid of 0\farcs20\,pixel$^{-1}$ using bilinear interpolation \citep[\texttt{reproject};][]{Robitaille2020}.

\section{Cumulative [O\,{\sc iii}] luminosity profile}
\label{sec:results_cumflux}

\begin{figure}[htb]
\begin{center}

    \includegraphics[width=0.9\linewidth]{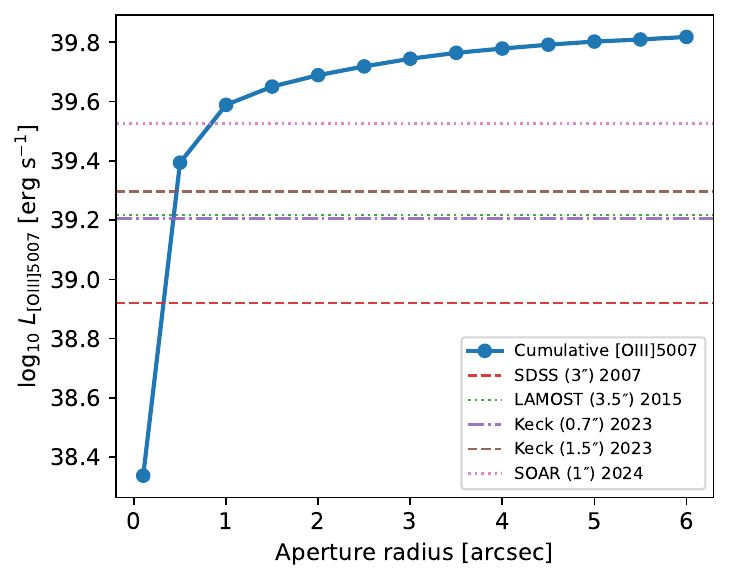}

    \caption{Cumulative [\ion{O}{III}]$\lambda$5007 luminosity as a function of aperture radius from the nucleus, derived from the MUSE WFM per-bin fluxes (blue solid line with points). Horizontal lines indicate [\ion{O}{III}]$\lambda$5007 luminosities presented in \cite{Sanchez-Saez24}, measured from archival single-aperture spectra at different epochs: SDSS ($3\arcsec$, 2007; red dashed), LAMOST ($3\farcs5$, 2015; green dotted), Keck ($0\farcs7$, 2023, purple dash-dotted; and $1\farcs5$, 2023, brown dashed), and SOAR  ($1\arcsec$, 2024; pink dotted).}
    \label{fig:oiii}
\end{center}
\end{figure}

Figure~\ref{fig:oiii} shows the cumulative [\ion{O}{III}]$\lambda$5007 luminosity as a function of aperture radius in the WFM field, integrating the per-bin luminosities from the nucleus outward. The profile rises steeply within the inner $\sim 1\arcsec$ and flattens at larger radii, indicating that the bulk of the total [\ion{O}{III}] luminosity is enclosed within a projected radius of $\sim 0.5$~kpc. The total integrated [\ion{O}{III}] luminosity within the WFM FoV is $L_{\rm [OIII]} = 5.5 \times 10^{39}$~erg~s$^{-1}$. As a reference, Figure~\ref{fig:oiii} shows the  $L_{\rm [OIII]}$ reported in \cite{Sanchez-Saez24}, where the authors claimed that the [\ion{O}{III}] flux evolved after $\sim 3.6$ years after the first ZTF alert. The VLT/MUSE observation thus confirms that the reported evolution of the $L_{\rm [OIII]}$ was real and not due to aperture effects.  

\section{Dust-reddening estimates from Balmer decrement}\label{app:red_map}

\begin{figure}
  \centering
  \includegraphics[width=0.9\linewidth]{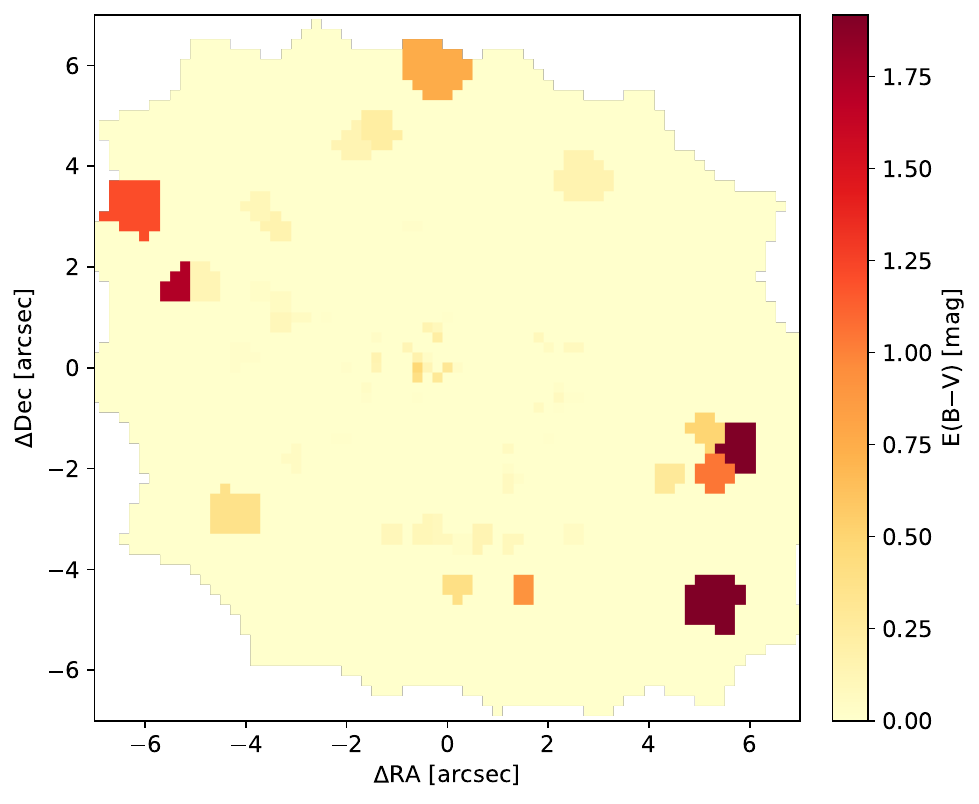}
  \caption{Dust-reddening map $E(B-V)$ of SDSS1335+0728 derived from
    the Balmer decrement in the MUSE WFM data. Coordinates are offsets from the nucleus in
    arcseconds.}
  \label{fig:ebv}
\end{figure}

The colour excess $E(B-V)$ is estimated on a bin-by-bin basis from the Balmer decrement. We adopt an intrinsic $\mathrm{H}\alpha/\mathrm{H}\beta$ ratio of 2.86, appropriate for Case~B recombination \citep[e.g.][]{Osterbrock2006}, and the \citet{Cardelli89} extinction law with $R_V = 3.1$. For each spatial bin with positive $\mathrm{H}\alpha$ and $\mathrm{H}\beta$ fluxes and an observed ratio exceeding the intrinsic value, the
reddening is obtained as $$E(B-V) = \frac{2.5}{k(\mathrm{H}\beta) - k(\mathrm{H}\alpha)} \log_{10}\!\left(\frac{F_{\mathrm{H}\alpha}/F_{\mathrm{H}\beta}}{2.86}\right),$$ where $k(\mathrm{H}\alpha) = 2.53$ and $k(\mathrm{H}\beta) = 3.61$ are the extinction coefficients at the respective wavelengths. Figure \ref{fig:ebv} shows the dust-reddening map $E(B-V)$ of SDSS1335+0728. The map shows negligible reddening overall ($E(B-V) \lesssim 0.1$~mag); the few bins with elevated values are likely driven by low-SNR $\mathrm{H}\beta$ detections.

\section{Minimum ionising luminosity from the H$\beta$ line}\label{app:lion_hbeta}

\begin{figure}[htbp]
\begin{center}
    \includegraphics[width=0.85\linewidth]{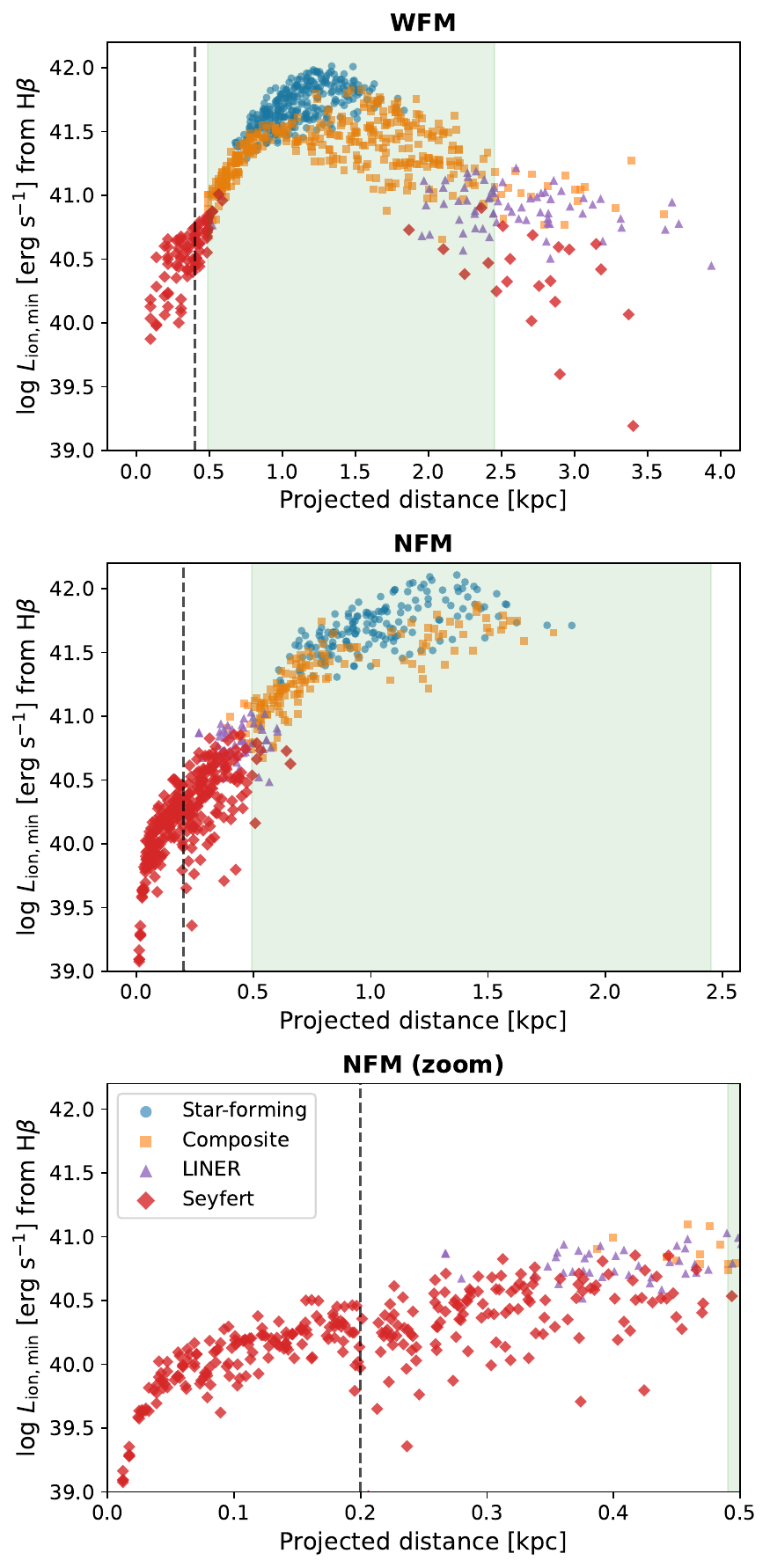}

\caption{As in Figure \ref{fig:muse_lion}, minimum ionising luminosity $L_{\rm ion,min}$ as a function of projected distance from the nucleus, but derived from the H$_{\beta}$ line.
\label{fig:muse_lion_hb}}
\end{center}

\end{figure}

Figure~\ref{fig:muse_lion_hb} presents the minimum ionising luminosity $L_{\rm ion,min}$ derived from the H$\beta$ line as a function of projected distance, complementing the H$\alpha$-based analysis discussed in Section~\ref{sec:results_lion}. The resulting radial profiles are fully consistent with the H$\alpha$-based results.

\section{Silicate feature strength determination}\label{app:silicate}

\begin{figure}[htbp]
\begin{center}
        \includegraphics[width=1\linewidth]{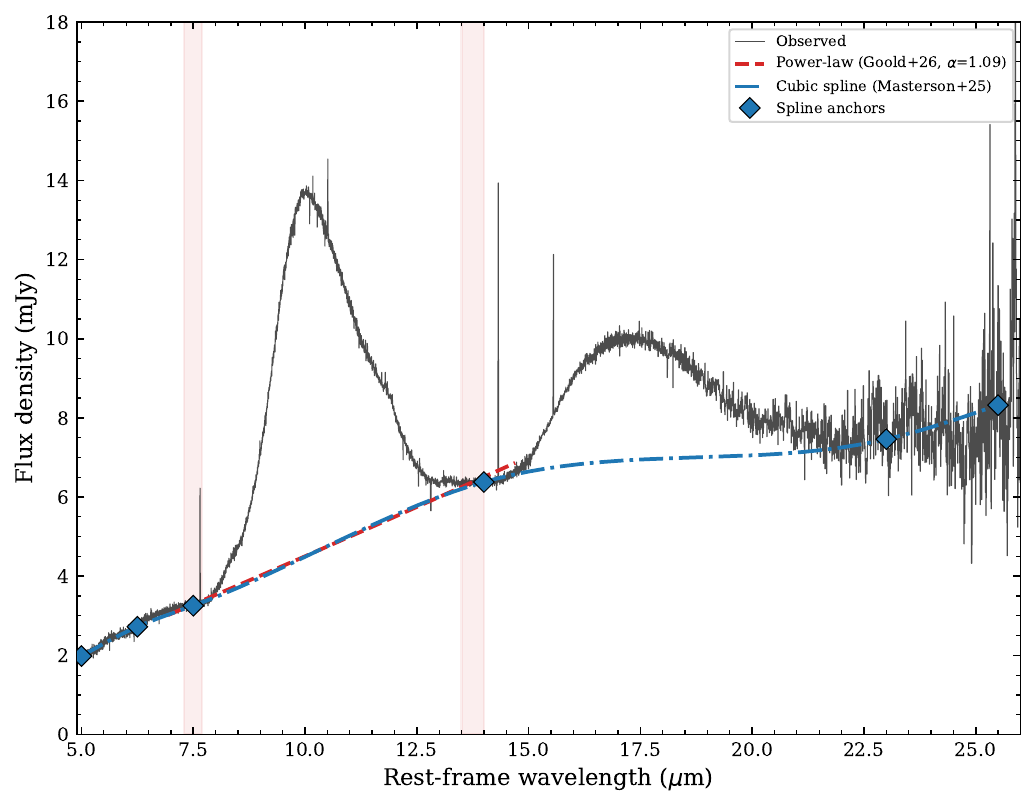}

    \caption{Measurement of the 9.7\,$\mu$m and 18\,$\mu$m silicate feature strengths in SDSS\,J1335$+$0728. The grey curve shows the observed MIRI/MRS spectrum (rest-frame). The red dashed line shows the power-law continuum fit of 
    \citet{Goold26} ($\alpha = 1.09$), anchored in the rest-frame windows 7.3--7.7 and 13.5--14.0\,$\mu$m (shaded regions). The blue dot-dashed line shows the natural cubic spline continuum of \citet{Masterson25}, with anchor points indicated by blue diamonds at 5.0, 6.25, 7.5, 14.0, 23.0, and 25.5\,$\mu$m.}
    \label{fig:jwst_silicate}
\end{center}
\end{figure}

We measure the silicate feature strength from the 1D JWST MIRI/MRS spectrum using two independent continuum estimation methods shown in Figure \ref{fig:jwst_silicate}. Following \cite{Goold26}, we fit a power-law continuum anchored at 7.3--7.7 and 13.5--14.0\,$\mu$m (rest-frame) and define the silicate strength as $S_{\rm sil} = \ln(f_{\rm obs}/f_{\rm cont})$ at 9.7\,$\mu$m ($S_{\rm sil}$[9.7$\mu$m]), adopting the original \cite{Spoon07} sign convention (note that \citealt{Goold26} followed the sign convention of \citealt{Spoon22}). Moreover, following \citet{Masterson25}, we alternatively fit a cubic spline through six rest-frame anchor points at 5.0, 6.25, 7.5, 14.0, 23.0, and 25.5\,$\mu$m and measured $S_{\rm sil}$ at 9.7\,$\mu$m and 18\,$\mu$m ($S_{\rm sil}$[18$\mu$m]). Following power-law continuum approach, we obtained $S_{\rm sil}$[9.7$\mu$m]~$=0.99$, while following the cubic spline continuum approach, we obtained $S_{\rm sil}$[9.7$\mu$m]~$=1.12$ and $S_{\rm sil}$[18$\mu$m]~$=0.36$. 

\end{appendix}

\end{document}